\documentclass[a4paper,12pt]{report}

\usepackage{tikz}
\usetikzlibrary{plotmarks,shapes,arrows,chains,hobby,backgrounds,calc,trees,decorations.pathreplacing}
\usepackage{arydshln}

\usepackage{amsmath, amssymb, amsthm} 
\usepackage{graphicx,color}           
\usepackage[left=1in, right=1in, top=1in, bottom=1.3in, includefoot, headheight=13.6pt]{geometry}

\usepackage[T1]{fontenc}
\usepackage{amscd}
\usepackage{amsfonts}
\usepackage{amsmath}
\usepackage{amssymb}
\usepackage{amstext}
\usepackage{amsthm}
\usepackage[titletoc]{appendix}
\usepackage{bm}
\usepackage{booktabs}
\usepackage{calligra}
\usepackage[justification=raggedright,singlelinecheck=false,font={small,it},labelfont={bf,it}]{caption}
\usepackage{enumerate}
\usepackage{epstopdf}
\usepackage{fancybox}
\usepackage{fancyhdr}
\usepackage{float}
\usepackage{graphicx, adjustbox}
\usepackage{listings}
\usepackage{longtable}
\usepackage{courier}
\usepackage{mathrsfs}
\usepackage[parfill]{parskip}[2001/04/09]
\usepackage{rotating,lscape}
\usepackage{subfigure}
\usepackage{verbatim}
\usepackage{footnote}
\makesavenoteenv{tabular}
\makesavenoteenv{table}
\usepackage[dvips]{epsfig}
\usepackage[dvips]{graphicx}
\usepackage{mathtools}

\usepackage[raggedright]{titlesec} 
\usepackage[flushleft]{threeparttable} 
\usepackage[justification=centering]{caption} 

\usepackage{setspace}
\usepackage{amsthm,amssymb,amsmath}

\newtheoremstyle{myplain}
  {\topsep}   
  {\parsep}   
  {\itshape\doublespacing}  
  {0pt}       
  {\bfseries} 
  {.}         
  {5pt plus 1pt minus 1pt} 
  {}       
\theoremstyle{myplain}

\newcommand{\rom}[1]{\uppercase\expandafter{\romannumeral #1\relax}}
\newcommand*{\Rom}[1]{\expandafter\@slowromancap\romannumeral #1@}
\floatstyle{plain}

\newcommand{\myqed}{\hskip 10pt $\square$}

\def\@submitdate{\ifcase\the\month\or
  January\or February\or March\or April\or May\or June\or
  July\or August\or September\or October\or November\or December\fi
  \space \number\the\year}

\usepackage{hyperref}
\hypersetup{
	colorlinks,
	citecolor=black,
	filecolor=black,
	linkcolor=black,
	urlcolor=black
}

\usepackage[bottom]{footmisc} 
\usepackage{natbib}
\usepackage{graphicx}
\usepackage[nottoc,notlot,notlof]{tocbibind}
\usepackage{relsize} 
\newcommand{\plus}{\scalebox{.5}{+}}
\newcommand{\minus}{\scalebox{.8}{-}}

\newcommand{\specialcell}[2][l]{%
  \begin{tabular}[#1]{@{}l@{}}#2\end{tabular}}

\newcommand{\mysc}[1]{\sl{\sc #1}} 

\fancypagestyle{empty}{ 
   \fancyhead{}
   
}

\begin{document}

\thispagestyle{empty}
\begin{titlepage}

\pagenumbering{gobble}
\begin{center}
\vspace{5cm}
\begin{table}
    \begin{tabular}{cc}

         & \\
    \end{tabular}
\end{table}

{\LARGE \bf Bayesian Modelling\\ { \large \bf of\\} \vspace{0.2cm}\LARGE Visual Discrimination Learning in Mice} \\
\vspace{1.5cm}
\href{mailto:pb736@cantab.ac.uk}{Pouya Baniasadi, PhD} \\ \vspace{0.15cm}
Department of Physiology, Development and Neuroscience \vspace{0.15cm}

\begin{center}
\includegraphics[scale=0.12]{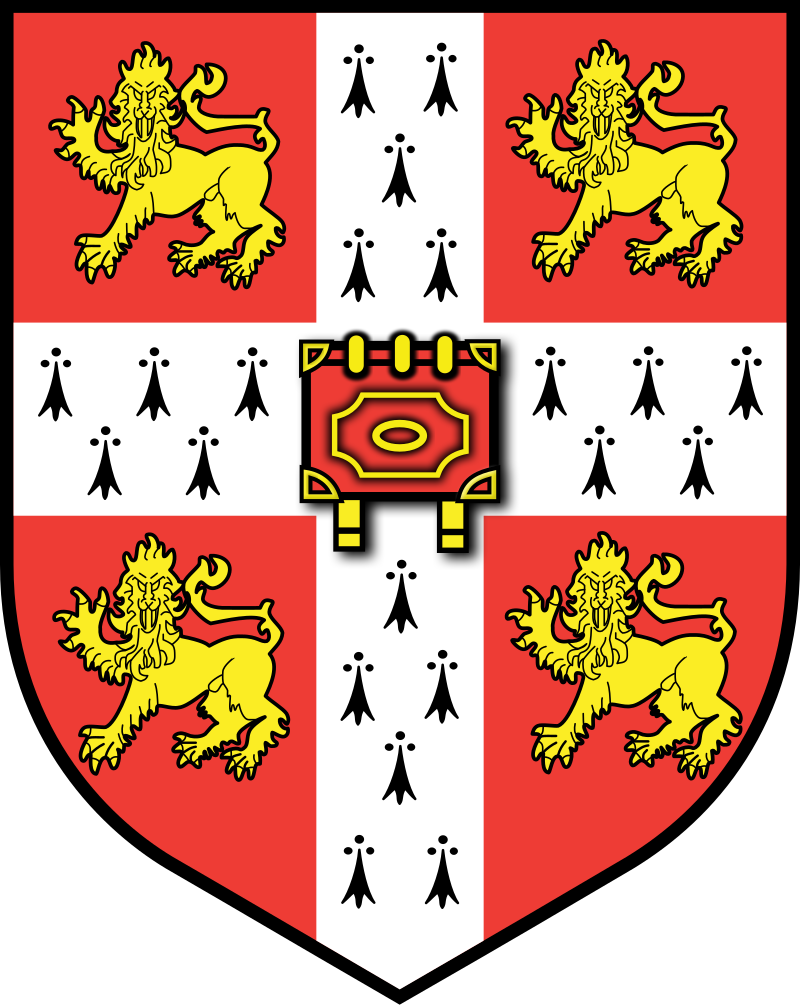} \\
\Large{\bf UNIVERSITY OF CAMBRIDGE}
\end{center}

\vspace{0.5cm}

August 2020\\ \vspace{0.6cm}
\textit{\small{ 
This project report is written in partial fulfilment of the requirement for the\\ 
Master of Philosophy in Basic and Translational Neuroscience}}\\
\end{center}

 \small{
\begin{table} [b!]
\begin{center}
\textbf{\it }\\
\begin{tabular}{ccc}
 \multicolumn{3}{c}{\textbf{\it Supervised by \ \ \ \ }} \\ \\ 
     \specialcell{Dr. Jasper Poort} & & \specialcell{Prof. M{\'a}t{\'e} Lengyel} \vspace{0.15cm }\\
     \specialcell{Selective Vision Laboratory}  &\hspace{2cm} & \specialcell{Computational and Biological \\ \ \ \ \ \ \ \ Learning Laboratory} \vspace{0.15cm }\\
     \specialcell{Department of Psychology} & & \specialcell{Department of Engineering}\\
\end{tabular} 
\end{center}
\end{table}

}

\end{titlepage}

\pagenumbering{roman}


\pagebreak

\chapter*{Dedication}

\vspace{3cm}
\begin{center}
\emph{For my parents Mahin and Ghasem, who taught me about pursuing dreams,\\ \vspace{0.25cm}
for their endless love, support and sacrifices}
\end{center}

\chapter*{Declaration}

\addcontentsline{toc}{chapter}{Declaration}
This report describes work carried out at Cambridge University from Jan 2020 to Jul 2020 under the supervision of Dr Jasper Poort (Selective Vision Laboratory at the Department of Psychology) and Prof. M{\'a}t{\'e} Lengyel (Computational and Biological Learning Lab at the Department of Engineering) as a part of the MPhil program in Basic and Translational Neuroscience. I confirm that the material in this report is not copied from any published material, nor is it a paraphrase or abstract of any published material unless it is identified as such and a full source reference is given. I confirm that, other than where indicated above, this document is my own work.
\vspace{1cm}
\begin{flushright}
\hspace*{8.5cm} \hrulefill \\[8pt]
Pouya Baniasadi\\
August 2020
\end{flushright}

\chapter*{Abstract}
\addcontentsline{toc}{chapter}{Abstract}
The brain constantly turns large flows of sensory information into selective representations of the environment. It, therefore, needs to learn to process those sensory inputs that are most relevant for behaviour. It is not well understood how learning changes neural circuits in visual and decision-making brain areas to adjust and improve its visually guided decision-making. To address this question, head-fixed mice were trained to move through virtual reality environments and learn visual discrimination while neural activity was recorded with two-photon calcium imaging. Previously, descriptive models of neuronal activity were fitted to the data, which was used to compare the activity of excitatory and different inhibitory cell types. However, the previous models did not take the internal representations and learning dynamics into account. Here, I present a framework to infer a model of internal representations that are used to generate the behaviour during the task. We model the learning process from untrained mice to trained mice within the normative framework of the ideal Bayesian observer and provide a Markov model for generating the movement and licking. The framework provides a space of models where a range of hypotheses about the internal representations could be compared for a given data set.

\pagebreak
\tableofcontents
\pagebreak

\pagenumbering{arabic}

\chapter{Introduction}

Learning modifies neural representations of behaviourally relevant information. While changes in response selectivity to behaviourally relevant stimuli have been observed in many studies across different species \citep{yang2004effect,yan2014perceptual,poort2015learning}. There has been growing evidence that different cell types, classified using molecular and cellular properties \citep{kepecs2014interneuron}, have specific roles in learning \citep{khan2018, fishell2019interneuron}. However, the nature of these changes and how they relate to sensory coding is not well understood \citep{yap2018activity}.

Probabilistic models of behavioural learning are an important approach to link the changes in neural representations to internal representation of the environment and decision-making \citep{fiser2010statistically,berkes2011spontaneous,heeger2017theory}. Given the non-deterministic nature of events in the real world, human and animal learning must involve at least some internal representations of the uncertainties in the environment \citep{barlow1961possible}. There has been an extensive body of research on how the nervous system represents uncertainly about the environment \citep{pouget2003inference, beck2008probabilistic, fiser2010statistically, kriegeskorte2018cognitive}. 

Bayesian learning theory provides a normative framework of learning that represents uncertainty in probabilistic outcomes\citep{bishop2006pattern}. In particular, the ideal observer analysis uses the Bayesian learning theory for achieving optimal learning performance in a given task \citep{geisler2003ideal,geisler2011contributions}. Learning can be conceptualised as the incorporation of sensory information to update and improve performance on a given task. the ideal observer performs at the theoretical limits of information processing to update their beliefs. 
However, it is important to note that optimality in this context refers to the optimal incorporation of information, which is not equivalent to achieving the optimal solution in all trials.
While the nervous system may or may not have representations similar to an ideal observer, the ideal observer analysis provides a systematic framework to formulate hypotheses about the internal representations and learning dynamics \citep{maloney2009bayesian,orban2008bayesian}.

In this thesis, we describe a Bayesian learning model using the framework of ideal observer learning. 
Our goal is to develop a model of internal representations of reward and space that are used for learning and adjusting behaviour in the visual discrimination task. This model will allow us in future work to relate the neuronal activity measurements \citep{poort2015learning,khan2018} to the internal representations that guide behaviour. We continue this chapter with a brief overview of the basic mathematical ideas used to develop the model. Then, in Chapter \ref{chap:experiment}, we explain the experimental setup and describe the behavioural data. A space of models (for the structure of Markov models) is introduced in Chapter \ref{chap:model_part1} which defines the internal representations of reward and state transitions. Then, a Bayesian model of learning reward probabilities and state transitions is described that uses the ideal observer framework. In Chapter \ref{chap:model_part2}, we introduce a generative Markov model that uses internal representations to generate behaviour. We also discuss the use of maximum likelihood estimation to estimate the model parameters. Finally, in Chapter \ref{chap:discussion}, we discuss the potential applications and limitations of the model and set out a path for the continuation of the research. 

\section{Mathematical preliminaries}

In this section, I briefly introduce the concepts that provide the mathematical foundation of the Behavioral model.

\subsubsection*{\emph{\textbf{Markov chain model}}}
A system has the \emph{Markov property} if the predictions about future events only require the knowledge of the system's present state. In other words, given the present state of the system, future events are conditionally independent of past events. A \emph{Markov chain} is a stochastic model of a sequence of events with the Markov property. 

\begin{figure}[ht]
\centering
\includegraphics[scale=0.4]{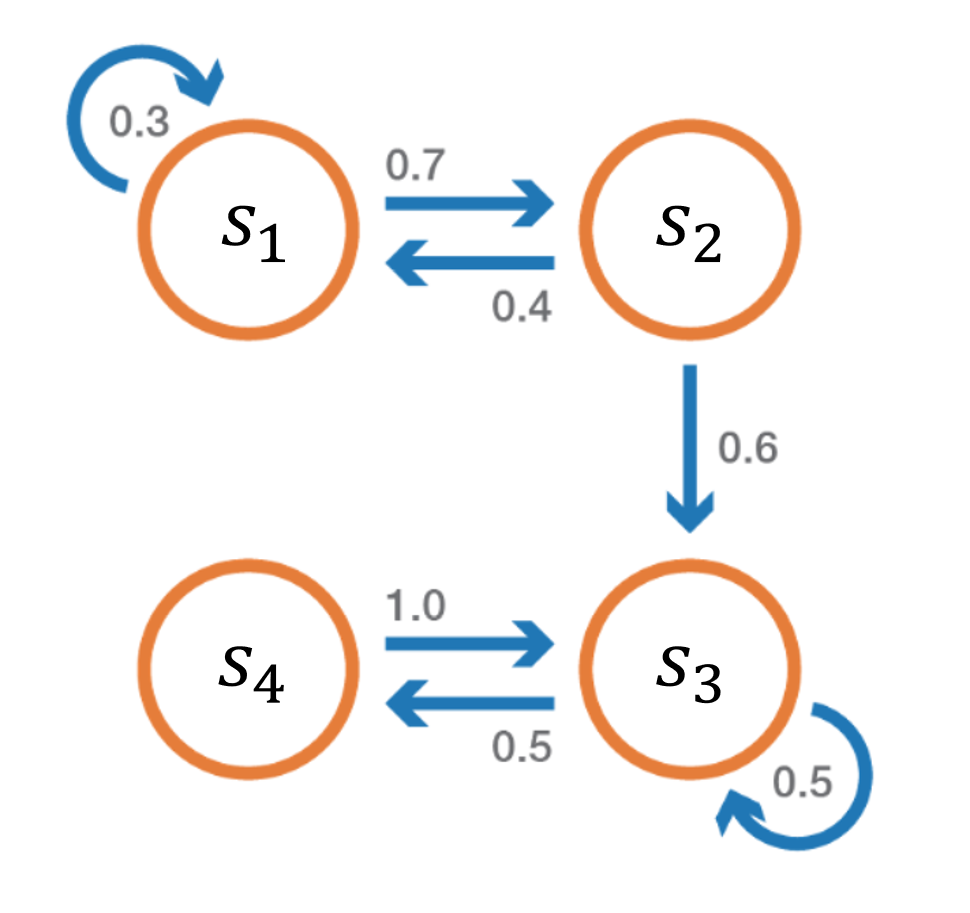}
\caption{Movement in a corridor simulated in the VR environment.}
\label{fig:MC_example}
\end{figure}

Let $S = {s_1, s_2,...,s_r}$ be a set of \emph{states} for a Markov chain. The process starts in one of these states and moves sequentially from one state to another. Each move is called a \emph{step}. Let $X_n$ be the current step. We denote by $p_{ij} = P(X_{n + 1} = s_j | X_{n} = s_i)$, the \emph{transition probability} of visiting state $s_j$ after visiting $s_i$. Note that by Markov property, given $X_{n}$, $X_{n + 1}$ is conditionally independent of the past states. A transition from $s_i$ to $s_j$ can be represented as a directed edge $(s_i, s_j)$ with a corresponding transition probability $p_{ij}$. The sum of transition probabilities of the outgoing edges from a state should add up to 1. Figure \ref{fig:MC_example} illustrates a Markov chain with 4 states and transition probabilities. 

Let $T = [p_{ij}]$ be the transition probability matrix for the Markov chain and let $\bm{u}$ be the
probability vector which represents the starting distribution (i.e., $X_k \sim \bf{u}$). Then the probability
that the chain is in state $s_i$ after $m$ steps is the $i$-th entry in the vector $\bm{u}(m) \coloneqq \bm{u}\ T^m$. That is,
\begin{align}\label{eq:MC_Pm+k}
P(X_{k+m}|X_{k}) = \bm{u}^{(i)}(m),\ \ \ \ \ \ \  \mbox{        where } \bm{u}(m)= \bm{u} T^m.
\end{align}

\subsubsection*{\emph{\textbf{Bayesian learning}}}

The probability of an event $A$ is denoted by $P(A)$. Consider another event $B$ and its corresponding probability $P(B)$. The conditional probability $P(A|B)$ is the conditional probability of $A$ given $B$. Bayes Theorem states that 
\begin{equation*}
    P(A|B) = \frac{P(B|A) ~ P(A)}{P(B)}
\end{equation*}

Consider a system that generates data and a space of possible models for describing the behaviour of the system. The probability distribution over the space of models $P(Model)$ represents the \emph{prior} knowledge about the system. Suppose that a set of data $D$ is observed from the system. Then $P(D~|~ Model)$ is called the \emph{likelihood} and $P(Data)$ is called the \emph{model evidence} or \emph{marginal likelihood}. 
The \emph{posterior distribution} over the models $P(Model~|~ D)$ represents our beliefs about the system after observing the data $D$. Bayes rule provides a principled way of updating our beliefs about the system after observing data. Formally, 
\begin{equation}\label{eq:BayesModel}
    P(Model ~|~ Data) = \frac{P(Data~|~Model) ~ P(Model)}{P(Data)}.
\end{equation}

\subsubsection*{\emph{\textbf{Dirichlet distribution learning of categorical probability values}}}

Consider a random variable which can take on $K$ possible categories. The \emph{categorical distribution} is a discrete probability distribution for the random variable, where the probability of each category is separately specified. The categorical distribution is a generalisation of the Bernoulli distribution for a discrete variable with more than two outcomes, such as the probability of outcomes for a 6-sided die. It is also a special case of the \emph{multinomial distribution} where the number of trials in one. 

If the probabilities of each outcome for a categorical distribution are unknown, using Bayesian learning, we can update prior probability distributions of probability values. 
The \emph{Dirichlet distribution} is a conjugate before the multinomial (and categorical) distribution, meaning starting with a Dirichlet prior and multinomial likelihood, the resulting posterior is also a Dirichlet distribution. The probability mass function for the Dirichlet distribution $Dir_K(\boldsymbol{\alpha})$ with $K$ categories is 
\begin{equation}
    f(\boldsymbol{p}| \boldsymbol{\alpha}) = \frac{1}{B(\boldsymbol{\alpha})}  \prod_{i=1}^{K} p_i^{\alpha^{(i)} - 1} , 
\end{equation}
where $ \boldsymbol{\alpha} = (\alpha^{(1)}, \ldots, \alpha^{(K)})$ is the vector of parameters. Furthermore, 
$${B(\boldsymbol{\alpha})} = \prod_{i=1}^{K} \frac{\Gamma(\alpha^{(i)})}{\Gamma\big(\sum^K_{i = 1} \alpha^{(i)}\big)}$$
where for positive real number $n$, $$\Gamma(n) = \int_{0}^{\infty}{x^{n-1} e^{-x} dx}.$$ For integer values, $\Gamma(n) = n!\ $.

To learn probabilities for a categorical distribution, given a prior distribution $Dir_K(\boldsymbol{\alpha})$ over the probability vector $\boldsymbol{p} = (p_1, \ldots, p_K)$, and data $\boldsymbol{c} = \{ c_1, \ldots, c_k \}$ representing the number of observation Dirichlet category, the posterior distribution is
\begin{align} \label{eq:dirch_post}
    P(\boldsymbol{p}|\boldsymbol{x}) & =  (\boldsymbol{p}|\boldsymbol{x} + \boldsymbol{\alpha}) \nonumber \\
    \boldsymbol{p}|\boldsymbol{x} & \sim Dir_K(\boldsymbol{x} + \boldsymbol{\alpha})
\end{align}

Finally, the Beta distribution is a special case of the Dirichlet distribution where the outcomes are binary (true or false). To distinguish this special case, we may use the notation $Beta(\beta^{(1)}, \beta^{(2)}) \equiv Dir_2(\boldsymbol{\alpha})$ where $\boldsymbol{\alpha} = \{\beta^{(1)}, \beta^{(2)}\}$.

\chapter{The experiment}\label{chap:experiment}
\vspace{-0.3cm}In this chapter, I describe the experimental setup in \cite{khan2018} and \cite{poort2015learning} for which we have developed a behavioural model in the later chapters. A summary of previous findings and a description of the behavioural data accompanied by figures are also included. 
\vspace{-0.3cm}\section{Experimental setup} \label{sec:exp_setup}
\begin{figure}[h!]
\centering
\includegraphics[scale=0.35]{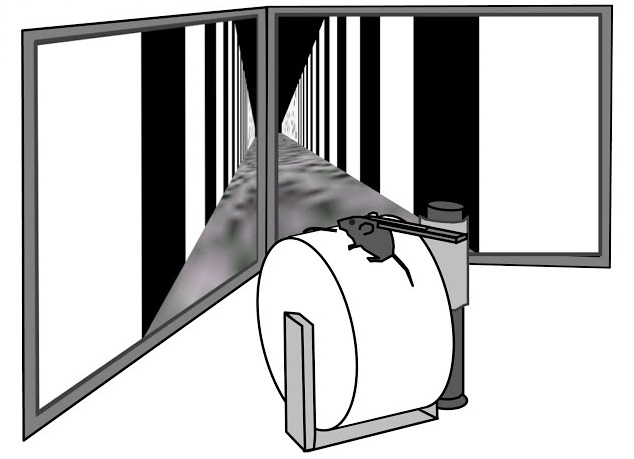}
\caption{Movement in a corridor simulated in the VR environment.}
\label{fig:VR_cylinder}
\end{figure}

The experimental setup involves the placement of the mouse on a cylindrical treadmill where its head is fixed to enable imaging of neural activity. The mouse can move forward (and backward). In front of the mouse, a screen is shown to the animal where visual feedback connected to the movement can simulate the movement of the subject in an environment. By controlling the setup of the space and visual stimulus while allowing imaging, the VR setup has been extensively used for studying the visual cortex and hippocampus in mice in recent years \citep{harvey2009intracellular,dombeck2010functional,khan2018,poort2015learning,saleem2018coherent}. Figure \ref{fig:VR_cylinder} illustrates the VR setup. 

\subsubsection*{\emph{Specifics of the corridor space and reward administration}}
\begin{figure}[h]
\centering
\includegraphics[scale=0.9]{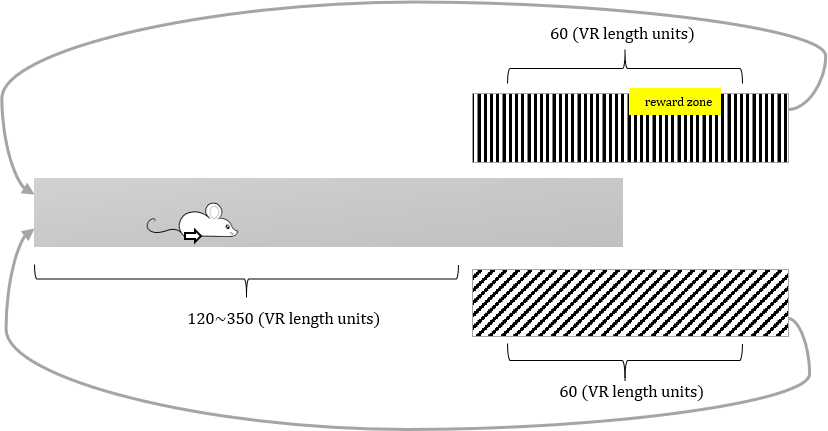}
\caption{Illustration of the corridor space.}
\label{fig:corridor_setup}
\end{figure}

We specifically consider the experimental setup described in \cite{khan2018, poort2015learning}. In these two studies, the activity of populations of neurons in V1 was measured with two-photon calcium imaging \cite{chen2013ultrasensitive} during a visual discrimination task in a virtual reality (VR) environment. Head-fixed mice ran through a simulated corridor where different types of visual stimuli were displayed on the walls. Three types of wall patterns characterise the different corridors. In the \emph{grey corridor} a short stretch of circle patterns followed by grey walls for a random distance, before the pattern on the walls abruptly changes to one of the \emph{grating corridors}. The grating corridors either displayed \emph{vertical gratings} (illustrated in Figure \ref{fig:VR_cylinder}) or \emph{angled gratings} for a fixed length (60 VR length units), before the grey corridor. An illustration of the corridor space is displayed in Figure \ref{fig:corridor_setup}. 

A milk dispenser was placed in front of the mouse to administer rewards. Mice received a reward for licking the dispenser in a reward zone starting halfway in the vertical grating corridor and halfway for around 10 VR-length units. If the mouse licked the dispenser in the reward zone, it would trigger the opening of the reward valve and a drop of soy milk would appear at the dispenser. No punishment was given for licking in the corridors with grey and angled grating walls. All mice learnt to discriminate the two stimuli, starting at the chance performance (behavioural $d^\prime$ close to zero) and reaching the threshold criterion of $d^\prime > 2.0$ within 5-9 days.

\subsubsection*{summary of previous findings}

The motivation behind developing a behavioural model is to take advantage of the behavioural data for the future analysis of experiments similar to \cite{khan2018}. A summary of results in \cite{khan2018} is as follows. After learning the visual discrimination task, neurons showed increased stimulus selectivity for the angled and vertical gratings. Interestingly, this effect depended on the cell types. In particular, stimulus selectivity for populations of pyramidal cells (PYR) along with parvalbumin (PV), somatostatin (SOM), and vasoactive intestinal peptide-expressing (VIP) inhibitory interneurons in layer 2/3 (L2/3) of the primary visual cortex (V1) were compared. Selectivity was increased for PYR and PV cells. PV neurons became as selective as the PYR cells, and showed changes in functional interactions, particularly with PYR cells. On the other hand, SOM neurons became decorrelated from the network and PYR–SOM coupling before learning predicted selectivity increases in individual PYR cells. While SOM inhibition seemed to gate changes in selectivity, PV cells provided strong stimulus selective inhibition after learning. A multivariate autoregressive linear model (MVAR model) fitted the activity of the neurons, and further supported the statistical analysis results. However, the MVAR model arguably neglects potentially important information in the behavioural data. Even though speed is taken into account, its contribution to the behaviour of the MVAR model is negligible. Accordingly, one of the primary motivations of the behavioural model proposed in this report is potential improvements in the (MVAR model). This is discussed in more detail in Chapter \ref{chap:discussion}.

\section{Behavioral data and observations}\label{sec:data_description}

Behavioural data were collected during the experiment. The distance travelled from the onset of the current corridor and the corridor type (determined by the wall patterns) is continuously recorded. The observed variables of spatial location and visual stimuli at each time are marked by a pair $(x,y) \in (xLoc \times Cor)$, where $x$ is the distance travelled from the onset of the current corridor pattern, and $y$ is the corridor pattern. Set $xLoc = [0, max(x)]$ is an interval from $0$ to the maximal length of an interval $max(x)$ and set $Cor = \{ grey, vertical, angled \}$ is the set of corridor types. The speed of the subject at each time is also recorded. A list of licking times and valve opening times (indicating reward administration) is also given by the data. 

For the generative behavioural model in Chapter \ref{chap:model_part2}, we discretize the data into time intervals of $\Delta\tau$ seconds, each identified by an index $t \in \{1, 2, \ldots, N\}$. The value of $\Delta\tau$ determines the time resolution of the behavioural data. Since the imaging data of \cite{khan2018} is taken in $\frac{1}{8}$ second intervals, time resolutions lower than $\frac{1}{8}$ seconds are not useful. Higher time resolutions may be desirable because they will decrease the computational cost of the analysis, but the cost of losing time resolution must be discussed. However, unless explicitly discussed, we can assume $\Delta\tau = \frac{1}{8}$ for the data analysis. Table \ref{tab:discrete_behavioral_data} describes the notation used to describe the data. Note that some of the records are behavioural, while others specify the values that are observed by the subject.

\begin{table} [ht!]
\begin{center}
\captionsetup{justification=raggedright,singlelinecheck=false}
\caption{Behavioral and observational records for $t \in\{ 1, 2 \ldots, N\}$. }\label{tab:discrete_behavioral_data} 
\begin{tabular}{|c|c|l|} \hline
     \textbf{Data} & Type &\multicolumn{1}{|c|}{\textbf{Description}} \\\hline
    $x_t$ & Observation &\specialcell{$x_t$ is the true value of distance from the onset of the current \\ corridor at time step $t$. }  \\ \hline
    $y_t$ & Observation & \specialcell{$y_t \in Cor = \{ grey, vertical, angled \}$ is the true value of the \\corridor type, which determines the visual stimuli at time\\ step $t$. } \\ \hline
    $o_t$ & Observation &\specialcell{$o_t$ is a binary value for whether the reward valve has opened\\ during the time step. }  \\ \hline
    $v_t$ & Behavior &\specialcell{Speed (average) at time step $t$}  \\ \hline
    $l_t$ & Behavior & Number of licks at time step $t$ \\ \hline
\end{tabular}
\end{center}
\end{table}

\subsubsection*{Instance of data visualisations}

\begin{figure}[h!]
\centering
\includegraphics[scale=0.50]{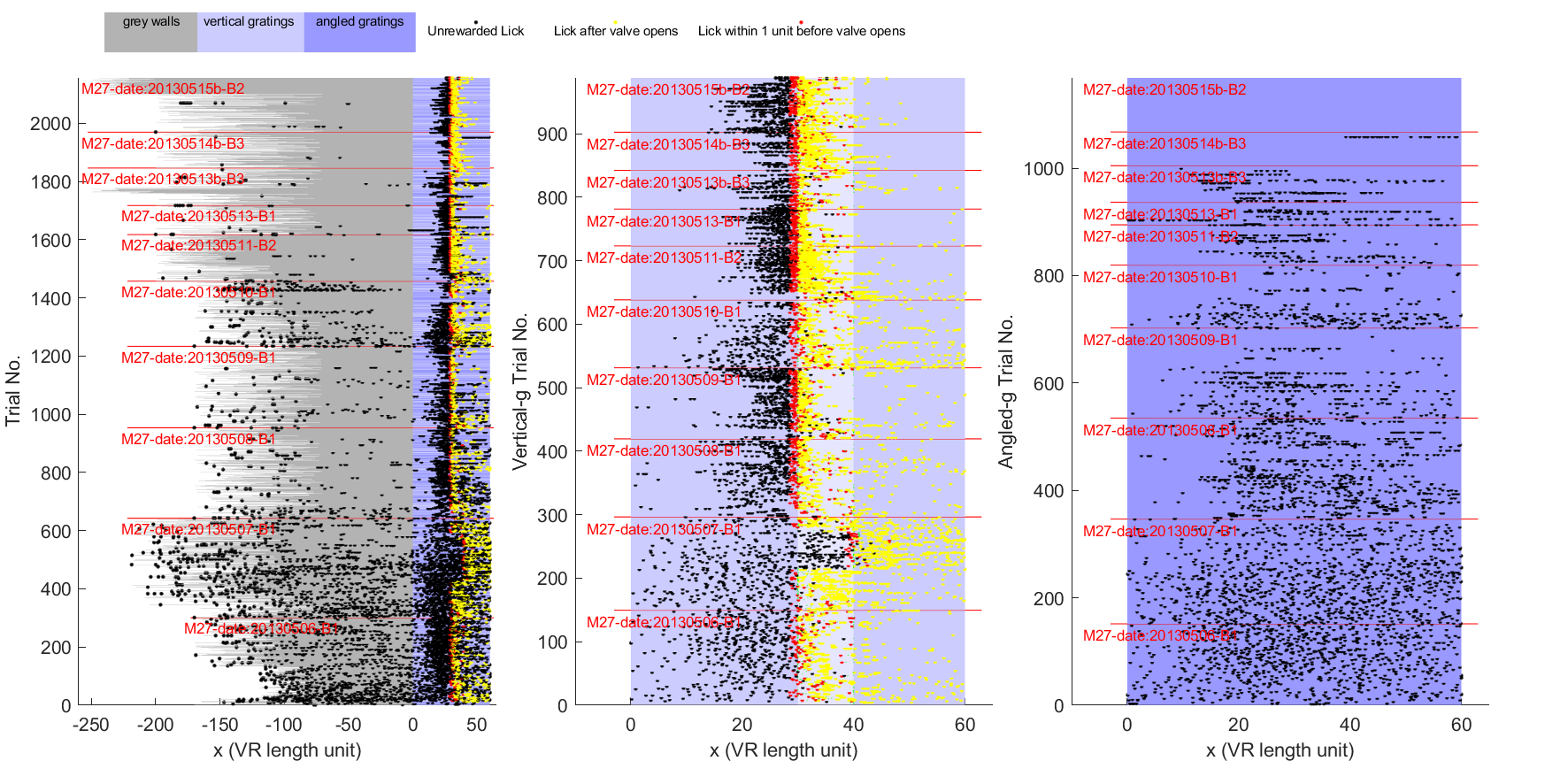}
\captionsetup{justification=raggedright,singlelinecheck=false}
\caption{Lick locations for M27. See the figure descriptions below. }\label{fig:licks_M27_main}
\vspace{1cm}
\includegraphics[scale=0.50]{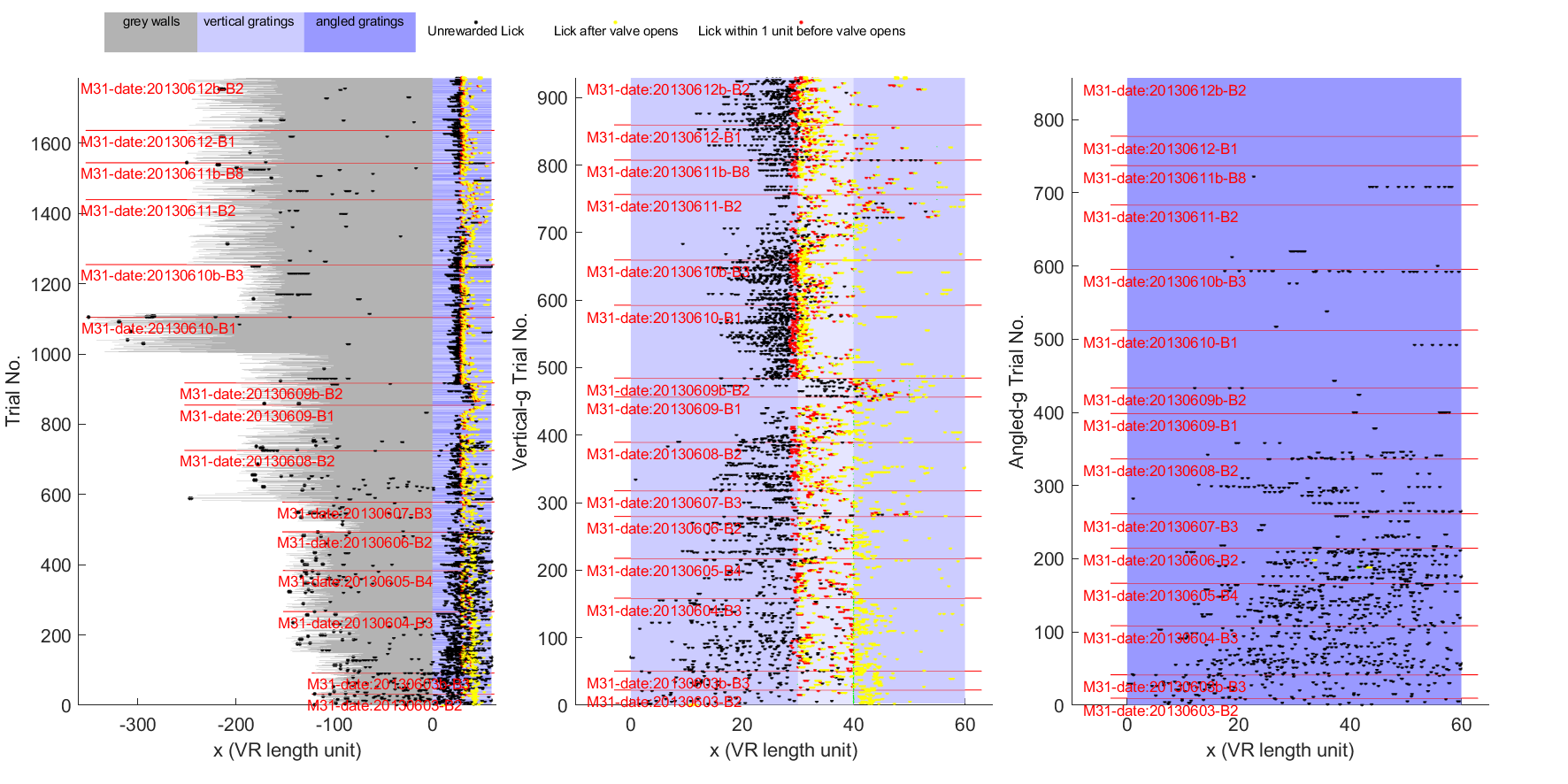}
\captionsetup{justification=raggedright,singlelinecheck=false}
\caption{Lick locations for M31 in all trials. The horizontal axis represents the location in a corridor, with 0 being set at the onset of a grating corridor. Negative values are in the grey corridors and positive values are in the grating corridors. The licking locations are marked by coloured points. Red dots represent licking within 1 length unit before a valve opening, and yellow indicates the licking after the opening of the reward valve, in a grating corridor. All other lick locations are marked in black. The trial number on the vertical axis shows the sequential order of the trials in each plot. The right plot shows all trials, where each trial is passing through one grey corridor followed by a grating corridor. The middle and the left plots show a closer look at the vertical and angled grating corridors. The red labels are labels for the experimental sessions.}
\label{fig:licks_M31_main}
\end{figure}

\begin{figure}[h!]
\centering
\includegraphics[scale=0.53]{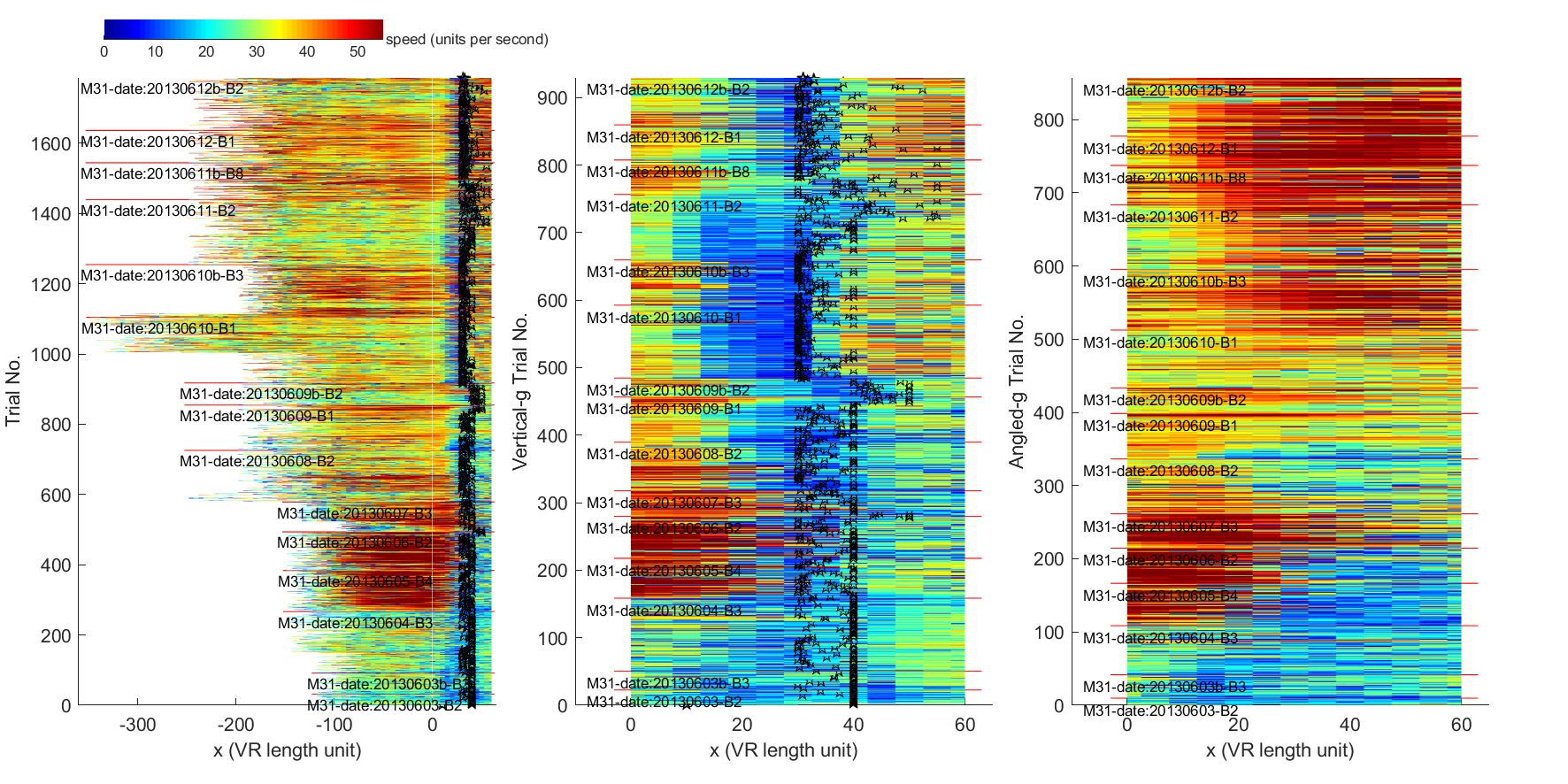}
\captionsetup{justification=raggedright,singlelinecheck=false}
\caption{Speed vs location for M31. See the figure descriptions below. }\label{fig:speed_M31_main}
\vspace{1cm}
\includegraphics[scale=0.53]{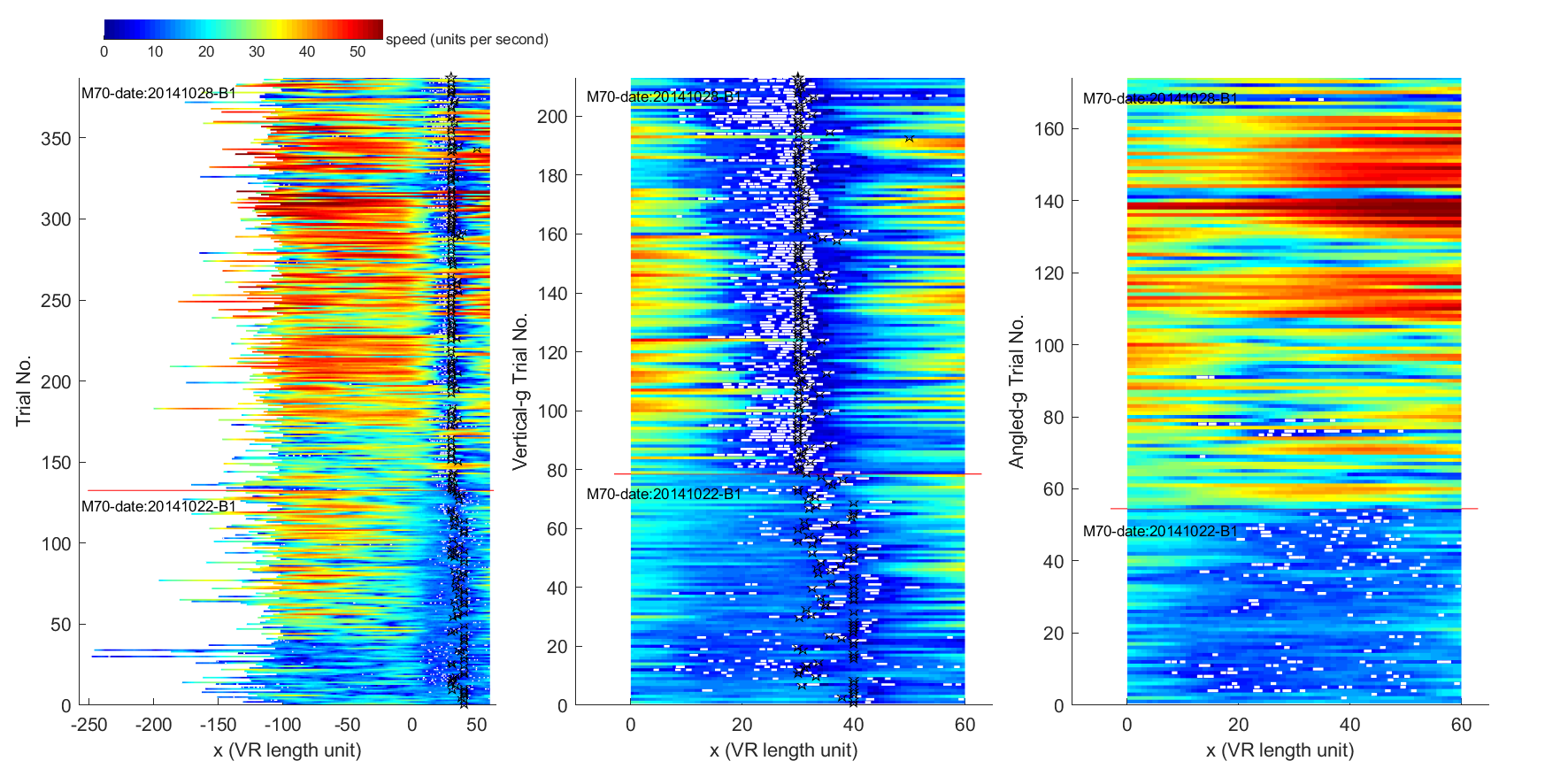}
\caption{
Speed and licks vs location for M70.  The horizontal axis represents the location in the corridor, with 0 being set at the onset of a grating corridor. Negative values are in the grey corridors and positive values are in the grating corridors. The trial number on the vertical axis shows the sequential order of the trials in each plot. The right plot shows all trials, where each trial is passing through one grey corridor followed by a grating corridor. The middle and the left plots show a closer look at the vertical and angled grating corridors. The colour for each location of each trial represents the speed of the animal at that point according to the colour scale; warmer colours represent higher speeds and cooler colours represent lower speeds. Note that for Figure \ref{fig:speed_M31_main}, the speed is averaged over 5 unit intervals due to virtual memory limits. The white points show the lock locations for M70, and the small black star indicates a valve opening location during a trial. The black labels are data labels associated with experimental sessions.
} \label{fig:speed_M70_main}
\end{figure}

The Figures below are instances of behavioural data visualizations from the experimental data. Figures \ref{fig:licks_M31_main} and \ref{fig:licks_M27_main} \citep{poort2015learning} illustrate the licking behaviour at different positions in different corridors, and Figures \ref{fig:speed_M31_main} \citep{poort2015learning} and \ref{fig:speed_M70_main} \citep{khan2018} give a colour map of speed at the different positions in the different corridors. For all Figures, the horizontal axis represents the position concerning the onset of the grating corridor \footnote{note that for the grey corridor this is obtained by shifting $x_t$ by the length of the grey corridor}, and the vertical axis is the trial index. Higher trial numbers are later. The black or red labels are data labels the ls associated with the experimental sessions.

The following observations about the licking behaviour have influenced parameter definitions and assumptions about prior beliefs of the animal in Chapter \ref{chap:model_part1}. These observations are consistent among all subjects. 
\begin{itemize}
    \item[] \textbf{reward association prior:} The mice do not know reward associations before the reward. However, the mice know that moving forward and licking the dispenser may lead to a reward. Initially, the licking behaviour is frequent to explore the space and discover reward associations. A uniformly random prior for reward probability may be appropriate. 
   \item[] \textbf{Change of visual discrimination:} The behaviour of the mice in the grating area and the grey area starts to diverge immediately, and the behaviour of the mouse in angled and vertical grating corridors seems to be similar at first; the differences of licking behaviour seem to be only after the reward is present in the vertical grating corridor. The dissociation of the award from the angled grating is realised substantially later than the dissociation of the reward from the grey area. It seems that at different points in the trial, the set of visually discriminated stimuli is different. 
    \item[] \textbf{Location is also taken into account} As the learning progresses, the licking concentrates close to the reward zone. It seems that the mice associate a spatial region, characterised by both visual stimuli and spatial positioning, with the reward area. 
\end{itemize}

The following observations about the speed have influenced our generative model of speed in Chapter \ref{sec:generativeModel}. These observations are consistent among all subjects. 
\begin{itemize}
    \item[] \textbf{Reward association influences speed}: the graphs suggest that the dissociation of reward in upcoming regions is associated with higher speed while anticipation of reward in upcoming regions is associated with reduction of speed. 
    \item[] \textbf{Evidence for change in the internal model:} while speed behaviour in the grey corridor diverges from the grating corridor quickly, the divergence of speed behaviour for angled grating and angled grating happen at a later point. This suggests that the mice initially correlate the grating areas with the reward, and then learn to differentiate between the grating areas to dissociate the angled grating with the reward. 
    \item[] \textbf{Change of visual discrimination::} Similar to the licking behaviour, initially speed behaviour seems to discriminate between the angled and vertical gratings only after the reward is present in the vertical grating corridor. This suggests that the mice initially correlate the grating areas with reward, and then learn to discriminate between the vertical and angled grating areas. 
\end{itemize}

\chapter{Behavioral model part 1: internal representations}\label{chap:model_part1}

\begin{figure}[b!]
\centering
\includegraphics[scale=0.8]{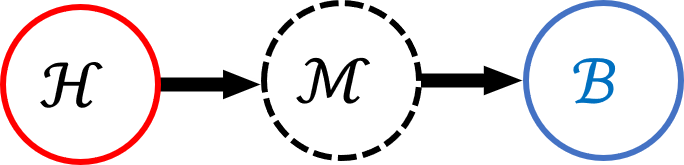}
\captionsetup{justification=raggedright,singlelinecheck=false}
\caption{Relation between history of experimental observations $\mathcal{H}$, internal model $\mathcal{M}$, and behavior $\mathcal{B}$. $\mathcal{H}$ and $\mathcal{B}$ are observed in the experimental data, but the internal model $\mathcal{M}$ is unobserved.}
\label{fig:GraphicalModel}
\end{figure}

\vspace{-1cm}
The behavioural model presented here provides a framework for inferring an internal model that can predict the animal's behaviour at a given time. Before getting into the specifics, consider a broad perspective on inferring a model that generates the current behaviour by incorporating past experiences. Figure \ref{fig:GraphicalModel} is a graphical model of big-picture relation between the history of animal's observations $\mathcal{H}$, the internal model $\mathcal{M}$ that incorporates experience into internal representations, and the observed behaviour $\mathcal{B}$. This chapter discusses the relationship between the history of observations and behaviorally relevant representations in the internal model ($\mathcal{H} \to \mathcal{M}$ in the graphical model of Figure \ref{fig:GraphicalModel}). I introduce a space of models where a range of hypotheses about the internal model can be systematically examined. The internal representations about reward and space are then used in the next chapter to construct a generative model of behaviour ($\mathcal{M} \to \mathcal{B}$ in the graphical model of Figure \ref{fig:GraphicalModel}). Then using a systematic approach, an internal model is inferred that best describes the data ($\mathcal{H}$ and $\mathcal{B}$). 

\pagebreak
By exploring and experiencing the environment, the brain uses experience to update its beliefs (i.e., learning) about the environment using its internal representations. In this learning model, the normative framework of Bayesian ideal observer analysis \citep{geisler2003ideal,geisler2011contributions} is used to learn behaviorally relevant internal representations. These include learning about the probability of reward in different regions of the VR corridor, and expectations about upcoming spatial regions when moving forward\footnote{The subject can only move forward due to the experimental setup.}. 

Model of spatial states in Section \ref{sec:state_model} describes how the space (VR corridor) is divided into states corresponding to spatial segments, where the representation of reward probability within a state only depends on the information (history of reward outcomes) obtained at that state. The structure of these states is a Markov chain. The space of models in Section \ref{sec:model_space} prescribes a range of Markov chain structures of spatial states within which a model is selected. For given states of a model, the dynamics for learning reward associations and state transitions are considered within the normative framework of the Bayesian ideal observer model in Section \ref{sec:learning_model}. 

\section{Structure of spatial states} \label{sec:state_model}

Animals' observation of visual stimuli and spatial positioning is an observation of the current $(x,y) \in \{xLoc, Cor\}$. Observations about reward association at the current location $(x,y)$ may be relevant to reward association at some other locations. It is therefore necessary to define spatial regions where reward observations are relevant to the entire region but explicitly irrelevant to other regions. To formalise this concept, the objective of this section is to associate the segments of space with states where the information about reward association is relevant to the current state and no other state. A reasonable way to define such states is to group areas that are spatially close by, visually similar, or both. 

\subsubsection*{\emph{\textbf{Defining states associated with spatial segments}}}
\begin{figure}[b]
\centering
\includegraphics[scale=0.9]{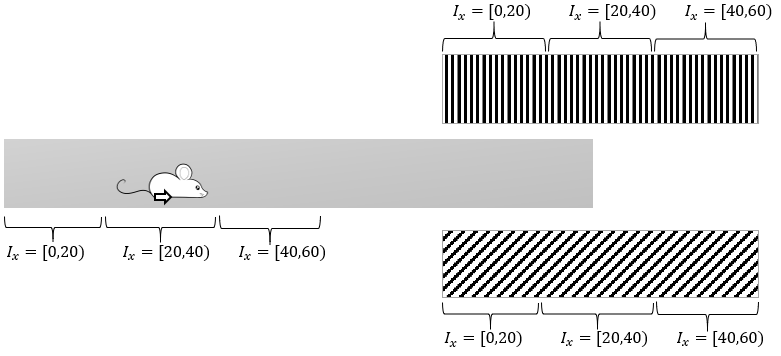}
\captionsetup{justification=raggedright,singlelinecheck=false}
\caption{An example of dividing the corridor space into mutually exclusive spatial segments. Each segment is then associated with exactly one state.}
\label{fig:corridor_segments}
\end{figure}

Taking into account both spatial proximity and visual similarity, consider sectioning $xLoc$ into a finite set of mutually exclusive spatial segments each identified by a fixed $y$, and an interval $I_x$ for $x$ values. We illustrate an example of spatial segmentation in Figure \ref{fig:corridor_segments}. Denote by $S$ a set of states and associate each segment with only one state (note that multiple segments may be associated with the same state). Then we say that \emph{the mouse is in state $s$} if its position $(x,y)$ is inside a segment that is associated with $s$. We associate all positions in all corridors with only one state with the function $f: (xLoc \times Cor) \to S$. The mouse may map locations onto states in multiple ways. By considering various ways to map between locations and states, we can infer the mapping that best matches the behavioural data (see \ref{sec:parameters_estimation}). 

\subsubsection*{\emph{\textbf{Spatial state transition event and structural properties}}}
Let $X_k$ be the random variable describing the $k$-th visited spatial state, where a spatial state transition event (i.e., transition to the next spatial step) happens when the subject \emph{crosses the initial point of a segment associated with a state}\footnote{Note that the time spent in each state is not fixed in this Markov model.}. Given the current position, the future positions do not depend on the history of visited positions, so given $X_k$, state $X_{k+1} $ is conditionally independent of $X_n$ for $n < k$. It follows that the state structure as defined above satisfies the Markov property. 

We assume that the spatial states are \emph{fully observable}. In other words, given a state structure, we assume that the subject always knows which state is the current state. Observations of the animal may be noisy and inaccurate, so assuming fully observable states is a simplification that may be contended with in a more sophisticated future model. However, states are associated with intervals of space rather than precise points in space, and they already incorporate some approximation about the spatial awareness of the subject. 

We assume that the mouse learns two things from the visual stimuli and licking in state $s$. First, it learns the reward association in that state. Second, it learns the transition from that state to other states. Let $r(s)$ be the probability that licking in state $s$ leads to reward in state $s$. Also, denote by $p_{(s,s^\prime)} = P(X_{k+1} = s^\prime | X_{k} = s)$ the transition probability of visiting any state $s^\prime$ after $s$. These parameters are initially unknown to the mouse and should be learned. In Section \ref{sec:learning_model}, I discuss a semi-normative model of learning for these parameters using the ideal observer framework. 

It is worth noting that the state transitions of the Markov chain are sparse. To understand the sparsity of state transitions, first note that $x$ is a positive real value, which ranges from $0$ to the maximal length of a corridor with the same patterns, and $y$ is a discrete value with three possible entries. From the onset of a corridor, until the onset of the next corridor, the spatial location is a continuous function of time. Within the period between two consecutive onsets, if a state transition happens, it can only be to the state associated with the next interval of $x$, with the same $y$. Moreover, when passing the onset of the next corridor, there is a discrete change in the value of $y$, and $x = 0$ at the onset of the new corridor. This event can only be a state transition to the start of a new corridor (a state that starts at $x = 0$) so there are at most three such possible transitions. It follows that the structure of states is a sparse Markov chain.

\section{Space of models for spatial states}\label{sec:model_space}

To define a space of models $\mathscr{M}$, we use two parameters for identifying a model in the model space; one for the set of discriminated patterns ($\mathcal{V}$), and one for the length of segments ($d$).

\subsubsection*{\emph{\textbf{Spatial model parameter $\mathcal{V}$: set of discriminated visual stimuli}}}
Let $\mathcal{V}$ be the set of visual stimuli that are discriminated in the spatial state model. The set of possible choices for $\mathcal{V}$ is $\{V_1, V_2, V_3 \}$ which are described below.
\begin{itemize}
\vspace{-.3cm}\item $V_1 = \{u\coloneqq\mbox{undifferentiated}\}$, where the grey and grating are not discriminated.
\vspace{-.5cm}\item $V_2 = \{g\coloneqq\mbox{grey}, va\coloneqq\mbox{angled or vertical grating}\}$, where the grey corridor is discriminated from the grating corridors, but where angled and vertical grating corridors are not discriminated.
\vspace{-.5cm}\item $V_3 = \{g \coloneqq \mbox{grey}, v \coloneqq \mbox{vertical}, a \coloneqq \mbox{angled}\}$, where the grey corridor, the angled and vertical grating corridor are discriminated.
\end{itemize}
While set $Cor$ contains the types of visual stimuli on the corridors, set $\mathcal{V}$ refers to subjective visual discrimination (or classification) between corridors by the mouse. Also note that the choices for set $\mathcal{V}$ implicitly contain a mapping from $Cor$ to $\mathcal{V}$.  

\subsubsection*{\emph{\textbf{Spatial model parameter $d$: length of states}}}
Denote by $d$ a value in the interval $(0, max(x)]$ for the length of spatial segments.
Value $d$ uniquely defines a sequence of intervals of $x$ values. For example, the associated sequence of intervals to $d = 30$ is $\{ [0, 30), [30, 60), \ldots\}$. Then state $s_{ij}$ is associated with the $j$-th interval of $x$, which is $[(j-1)d~,~ jd)$, and $i \in C$ identifies the visual stimuli. For example, for $\mathcal{V}=\{g, p\}$ and $d = 30$, the state $s_{p,2}$ refers to intervals of $x \in [30, 60)$ for both the vertical and angled grating corridors.

\begin{figure}[b!]
\centering
\includegraphics[scale=0.8]{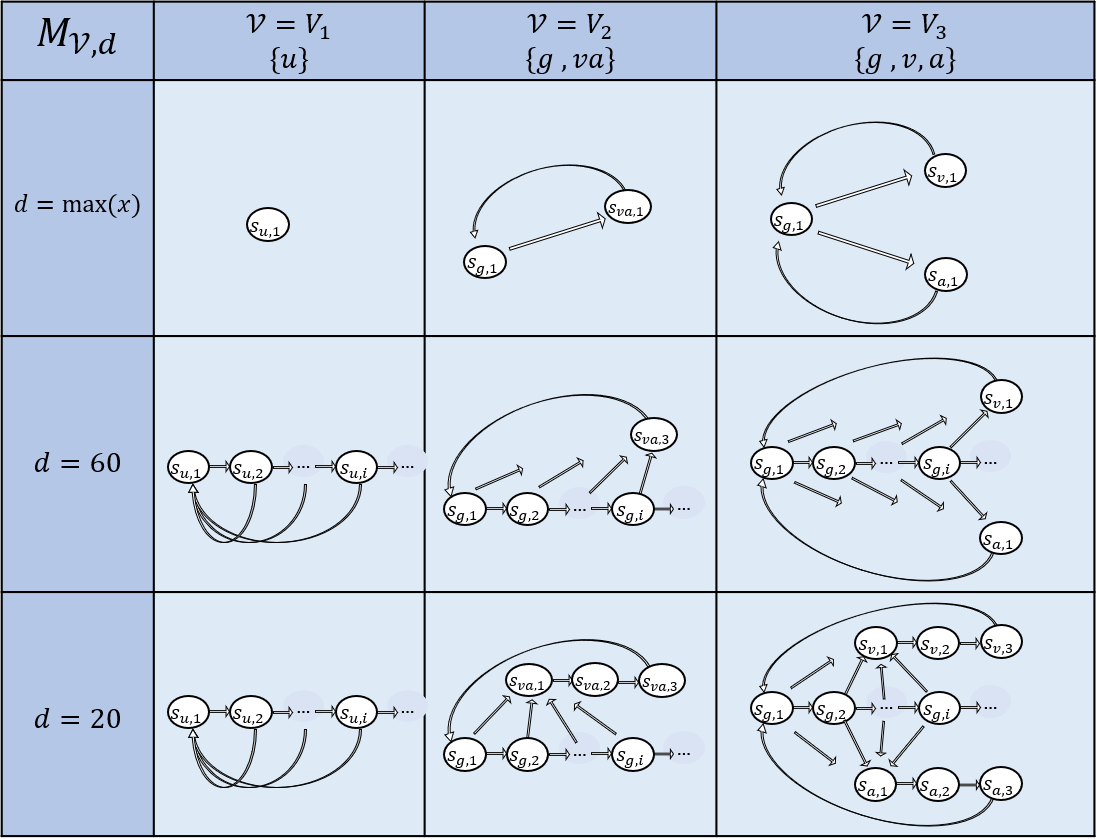}
\captionsetup{justification=raggedright,singlelinecheck=false}
\caption{Nine instances of Markov chain models $M_{\mathcal{V},d}$ for choices of $\mathcal{V}$ selected instances of $d$. For $d = x_max$, there is only one state per and self transition event only occurs when the corridor type changes. The length of the angled and vertically grating corridors is exactly 60 (VR length units) in the experiment. So for $d = 60$ and $d = 20$, there are exactly 1 and 3 states associated with the relevant element in $\mathcal{V}$. Note that the figure illustrates only selected instances of the model space $\mathscr{M}$. }
\label{fig:spaceofmodels}
\end{figure}

\subsubsection*{\emph{\textbf{Model space}}}
Now it is possible to introduce a Markov model $M_{\mathcal{V}, d} \in \mathscr{M}$ with the set of states $S$ that are associated with the spatial intervals induced by $\mathcal{V}$ and $d$. Since the length of the a corridor is bounded by $max(x)$, model $M_{\mathcal{V}, d}$ is a finite state Markov model. For example, $M_{V_1, max(x)}$ and $M_{V_3, max(x)}$ have exactly one and three states, respectively. Figure \ref{fig:spaceofmodels} illustrates the states of Markov chain models $M_{\mathcal{V}, D}$ for example cases of $\mathcal{V}$ and $d$.

Parameters $\mathcal{V}$ and $d$ are free parameters that will be set during the model selection, which will be further discussed in Section \ref{sec:parameters_estimation}. The fit for parameter $\mathcal{V}$, selected from $V_1$, $V_2$ or $V_3$, is determined by which stimuli the animal discriminates. The true value for $d$ is the length of spatial segments where information about reward associations and state transitions in the current segment is reasonably independent of segments associated with other states. For the sake of simplicity, it is assumed that $d$ is a fixed value, and it is the same across different visual stimuli. However, relaxing this assumption is possible by having more free parameters, for example, by introducing a free parameter of distance for each element of $\mathcal{V}$. For example, suppose $V = V_3$. Then instead of a free parameter $d$, we could use three parameters in $D = \{d_g, d_a, d_v\}$ which contains one free parameter of distance for every element of $\mathcal{V}$. In the initial implementation of the model, one parameter $d$ is considered. 

\begin{table} [t]
\begin{center}
\caption{Parameters for the model of spatial states} \label{tab:model_learning}.
\begin{tabular}{|c|c|l|} \hline
     \textbf{Parameter} & \textbf{Type} & \textbf{Description}  \\\hline
     $\mathcal{V}$ & \specialcell{Spatial model\\ parameter} & \specialcell{Set of discriminated visual stimuli on the corridors in the\\ model $M_{\mathcal{V},d}$; Possible options are $V_1 = \{u\}$, $V_2 = \{g,p\}$\\ and $V_3 = \{g,v,a\}$.}  \\\hline
     $d$ & \specialcell{Spatial model\\ parameter} &\specialcell{A constant length in $(0, max(x)]$ for the length of the\\ spatial for model $M_{\mathcal{V},d}$. }   \\\hline
\end{tabular} 
\end{center}
\end{table}

In summary, parameters $\mathcal{V}$ and $d$ for a model $M_{\mathcal{V}, d}$ determine the structure of the states in the Markov chain, where for each state the learning dynamics about reward association and state transitions is only dependent on the observations in that state. The learning dynamics are discussed in the next section. 


\section{Bayesian learning model}\label{sec:learning_model}

As first noted in Section \ref{sec:state_model}, in any state $s$, the subject uses sensory information to learn $r(s)$, the probability that licking in $s$ leads to the administration of reward in $s$, or \emph{reward probability of $s$} for short. Furthermore, state transition probability $p_{(s,s^\prime)}$, which is the probability of visiting state $s^\prime$ after visiting $s$, is also unknown to the subject and it is learned. Here, we use the ideal observer framework \citep{geisler2003ideal} to develop a semi-normative model for learning both reward associations and state transitions. In this section, the learning dynamics are discussed for a given model $M \in \mathscr{M}$. Therefore, states $S$ and their associated spatial intervals are unambiguous. 

\subsection{Learning reward probability within a state}

Recall that reward is given to the subject immediately after the subject licks the dispenser in the reward zone (see Section \ref{sec:exp_setup} for details of the experimental setup). The reward is a fixed amount of milk administered via the dispenser. We noticed that even in trained animals, licking started before the reward zone (see example mice in Figures \ref{fig:licks_M27_main} and \ref{fig:licks_M31_main}). This suggests that the mouse associates an extended region with the reward delivery which starts before the reward zone set by the experimenters. 

\subsubsection*{\emph{\textbf{Reward outcome $\bm{R}_k$ of current spatial step $k$}}}
If the mouse licks the dispenser in state $s$, it collects some information about the unknown parameter $r(s)$. If the subject does not lick the dispenser, it obtains no information about $r(s)$. Let the random variable $\bm{R}_k = (R^{(T)}_k, R^{(F)}_k)$ be the \emph{reward outcome of spatial step $k$}, where $R^{(T)}_k$ counts the number of positive outcomes, and $R^{(F)}_k$ counts the number of negative outcomes in spatial step $k$. As a consequence of the experimental setup, the amount of reward and the frequency of licking in the experiment does not provide any additional information about a reward region. Furthermore, spatial states are defined to be regions where licking at different points within the region does not provide additional information about the reward. Therefore, each visit to a state provides only three possible reward outcomes:
\begin{itemize}
    \vspace{-.3cm}\item $\bm{R}_k = ( 1, 0 )$ for subject licking the dispenser in spatial step $k$ followed by reward becoming available in spatial step $k$,
    \vspace{-.3cm}\item $\bm{R}_k = ( 0, 1 )$ for subject licking the dispenser in spatial step $k$ followed by no reward in spatial step $k$, and
    \vspace{-.3cm}\item $\bm{R}_k = ( 0, 0 )$ for subject not licking the dispenser in spatial step $k$.
\end{itemize}

\subsubsection*{\emph{\textbf{Normative model for updating internal reward representations (Bayesian) }}}
Let us first discuss how an ideal observer updates its prior beliefs about $r(s)$ after visiting state $s$ in spatial step $k$. The ideal observer provides a theoretical upper limit of performance, given the collected data. It is therefore a normative framework for updating the beliefs about reward association. Let prior beliefs about $r(s)$ right before visiting spatial step $k$ be a $Beta$ distribution 
\begin{align*}
Beta(\beta^{(1)}_k(s), \beta^{(2)}_k(s))
\end{align*}
over the interval $[0,1]$. The reward outcome $\bm{R}_k = (R^{(T)}_k, R^{(F)}_k)$ is the data that is newly collected about the reward. By Equation \ref{eq:dirch_post}, the posterior is
\begin{align*}
    r(s)|\bm{R}_k \sim Beta(R^{(T)}_k + \beta^{(1)}_k(s), R^{(F)}_k + \beta^{(2)}_k(s)).
\end{align*}

\subsubsection*{\emph{\textbf{Reward learning rate $\eta_r$}}}
The above is a theoretical bound on learning from observations in state $s$, assuming a prior $Beta$ distribution over $[0,1]$ for the reward probability $r(s)$. Some mice learn faster than others, and all of them will perform no better than the ideal observer model above. To allow for individual differences, and different learning rates, we introduce a model parameter $\eta_r \in [0, 1]$, which dials the amount of data required for the same amount of learning as an ideal observer. The \emph{update rule} (i.e., posterior) is
\begin{align*}
r(s)|\bm{R}_k & \sim Beta(\eta_r R^{(T)}_k + \beta^{(1)}_k(s), \eta_r R^{(F)}_k + \beta^{(2)}_k(s)).
\end{align*}

To keep track of learning parameters, let $\mathbf{B}_k = \big\{\bm{\beta}_k(s) \coloneqq \big(\beta^{(1)}_k(s), \beta^{(2)}_k(s)\big): s \in S\big\}$ be the beta parameters for beliefs about reward probabilities of all states in spatial step $k$. Note that after visiting state $s$ in spatial step $k$, 
\begin{align}\label{eq:r_update_rule}
     &\bm{\beta}_{k+1}(s) = \eta_r \bm{R}_k + \bm{\beta}_{k}(s) &&\mbox{for $s = X_k$, and} & \\
     &\bm{\beta}_{k+1}(s^\prime) = \bm{\beta}_{k}(s) &&\mbox{for $s^\prime \not= X_k$. } &  \nonumber
\end{align}

Note that $\eta_r$ is defined to have the same value across all states.  If $\eta_r = 1$, the mice performs as well as the normative ideal observer, and if $\eta_r = 0$, the mouse never learns reward associations. For the values in between $0$ and $1$, the mouse requires extra data points for updating its beliefs to the same extent as an ideal observer model. The model parameter $\eta_r$ can be interpreted as the data efficiency of learning. It could be used to compare individual learning differences among subjects. Furthermore, it is interesting to assess whether differences of $\eta_r$ in individuals is predictive of comparative learning rates on other learning tasks. It also provides a qualitative way to assess the model. For example, if the value is unreasonably high, it may indicate a flaw in the state structure or an incorrect choice of prior. 

\begin{table} [t]
\begin{center}
\captionsetup{justification=raggedright,singlelinecheck=false}
\caption{Guide for variables (Var) and parameters (Par) relevant to internal reward representations.} \label{tab:r_learning}.
\begin{tabular}{|c|c|l|} \hline
     \textbf{Var/Par} & \textbf{Type} & \textbf{Description}  \\\hline
     $\bm{R}_k$ & observed & \specialcell{A binary pair representing the reward outcome of step $k$,\\ 
     $(1,0)$ lick and reward within step $k$\\
     $(0,1)$ lick but no reward within step $k$\\
     $(0,0)$ no lick within  step $k$}  \\\hline

     $\mathbf{B}(k)$ & inferred & \specialcell{List of $\big(\beta^{(1)}_k(s), \beta^{(2)}_k(s)\big)$, for all $s \in S$, where
     \\$Beta(\big(\beta^{(1)}_k(s), \beta^{(2)}_k(s)\big))$ represents the beliefs about\\ $r(s)$ at spatial step $k$.} \\\hline
     $\eta_r$ & model parameter &  \specialcell{A constant in the $[0,1]$ interval for learning rate of\\ reward association.} \\\hline
\end{tabular}
\end{center}
\end{table}

\subsubsection*{\emph{\textbf{Implementation notes}}}
To simplify model implementation, we can derive the posterior distribution at step $k$ by merely keeping a list record of the total count of positive and negative reward outcomes in state $s$. In particular, at step $k$, for state $s$, let $\bm{c}_k(s) = \Big(c_k^{(T)}(s), c_k^{(F)}(s)\Big)$ be the total count of positive and negative outcomes in state $s$, from step $1$ up to the start of step $k$. That is,
$$\bm{c}_k(s) = \sum^{k}_{\substack{n=1\\ X_n = s}} \bm{R}_k.$$
For current spatial state $k$, a list of numbers can store values of $\bm{c}_k(s)$. Assuming a uniform prior at the start of the experiment, or $\beta^{(1)}_1(s) = \beta^{(2)}_1(s) = 1$, the prior probability distribution of $r(s)$ at step $k$ is
\begin{align*} 
    &r(s) \sim Beta(\eta_r c_k^{(T)}(s) + 1, \eta_r c_k^{(F)}(s) + 1),
\end{align*}
for which,
\begin{align}\label{eq:r_stepk_rule}
   &\bm{\beta}_k(s) = \eta_r \bm{c}_k(s) + 1.
\end{align}

\subsection{Learning state transitions}

Learning dynamics for state transitions $p_{(s,s^\prime)}$ is defined similarly to the reward associations. Let $E$ be the set of transition edges (directed edges), and let $Adj(s) = \{ s^\prime: (s,s^\prime) \in E \}$ be the set of states which for $X_{k} = s$, outcome of $X_{k}$ is in $Adj(s)$. Therefore, transition probabilities from $s$, $P(X_{k+1}|X_{k} = s)$ is a distribution of outcomes over $Adj(s)$. Assuming fixed probability transitions, $P(X_{k+1}|X_{k} = s)$ can be represented by a list of probabilities $\bm{p}(s) \coloneqq \big(p_{(s,s^\prime)}: s^\prime \in Adj(s)\big)$. Note that if the subject is not familiar with the space, the true distribution is unknown, and the subject learns about these probabilities through experience. 

\subsubsection*{\emph{\textbf{Normative model for updating internal transition representations (Bayesian)}}}
Every time the subject leaves state $s$ and the next step is observed, one observation is made about the outcome of $X_{k+1}$ given $X_{k} = s$. Because the outcome is a multinomial random variable, where possible outcomes are states in $Adj(s)$, we use a Dirichlet prior distribution to represent uncertainties about $\bm{p}(s)$. Specifically, at spatial step $k$, 
\begin{align*}
    \bm{p}(s) \sim Dir\big(\bm{\alpha}_k(s)\big)
\end{align*}
where the list of parameters $\bm{\alpha}_k(s)$ contains an element corresponding to each possible outcome. In particular, 
$$\bm{\alpha}_k(s) = \big(\alpha_{k}(s,s^\prime): s^\prime \in Adj(s)\big).$$ 

Suppose $X_{k} = s$ and consider an ideal observer whose prior beliefs about $\bm{p}(s)$ at spatial step $k$ is described by $Dir(\bm{\alpha_k}(s))$. Also suppose, the ideal observer visits the next state and makes the observation $X_{k+1} = \breve{s}$. Then by Equation \ref{eq:dirch_post}, the posterior distribution is 
\begin{align*}
    \bm{p}(s)|(X_{k+1} = \breve{s}, X_{k} = s)  \sim Dir\big(\bm{\alpha}_{k+1}(s)\big)
\end{align*}
where any element $\alpha_{k}(s,s^\prime)$ of $\bm{\alpha}_{k+1}(s)$ is updated as follows:
\begin{align*}
     &\alpha_{k+1}(s,s^\prime) = 1 + \alpha_{k}(s,s^\prime) &&\mbox{for $s^\prime = \breve{s}$, and} & \nonumber\\
     &\alpha_{k+1}(s,s^\prime) = \alpha_{k}(s,s^\prime) &&\mbox{for $s^\prime \not= \breve{s}$. } &  
\end{align*}
Furthermore, for any other state $s^{\prime\prime} \not= s$, it is obvious that the beliefs are not updated, i.e., $\bm{\alpha}_{k+1}(s^{\prime\prime} \not= s) = \bm{\alpha}_{k}(s^{\prime\prime} \not= s)$.

\subsubsection*{\emph{\textbf{Reward learning rate $\eta_p$}}}

Similar to introducing a learning rate for learning reward association, we introduce a $\eta_{p} \in [0,1]$ to account for data inefficiency compared to the ideal observer. Denote by $\mathbf{A}_k$, the list of all learning parameters of state transition probabilities $\mathbf{A}_k = \big(\bm{\alpha}_{k}(s): s \in S \big)$. Now, the \emph{update rule} (posterior distribution) is
\begin{align*}
    \bm{p}(s)|(X_{k+1}, X_{k})  \sim Dir\big(\bm{\alpha}_{k+1}(s)\big)
\end{align*}
where any element $\alpha_{k}(s,s^\prime)$ of a list of parameters in $\mathbf{A}_k$ is updated as follows:
\begin{align}\label{eq:p_update_rule}
     &\alpha_{k+1}(s,s^\prime) = \eta_{p} + \alpha_{k}(s,s^\prime) &&\mbox{for $s = X_{k}$ and  $s^\prime = X_{k+1}$} & \\
     &\alpha_{k+1}(s,s^\prime) = \alpha_{k}(s,s^\prime) &&\mbox{otherwise.} & \nonumber 
\end{align}
\begin{table} [t]
\begin{center}
\captionsetup{justification=raggedright,singlelinecheck=false}
\caption{Parameter guide for learning transition probabilities} \label{tab:p_learning}.
\begin{tabular}{|c|c|l|} \hline
     \textbf{Parameter(s)} & \textbf{Type} & \textbf{Description}  \\\hline
     $(X_{k+1}| X_{k})$ & observed & \specialcell{Transition outcome from a given state $X_{k}$\\}  \\\hline
     $\mathbf{A}_k$ & inferred & \specialcell{List of $\bm{\alpha}_k(s)$, for all $s \in S$, where $Dir(\bm{\alpha}_k(s))$ \\ represents beliefs about $\bm{p}(s)$ at step $k$ (list of state \\ transition probabilities from $s$ to adjacent states)} \\\hline
     $\eta_p$ & free parameter &  \specialcell{A constant in the $[0,1]$ interval for learning rate of\\ transition probabilities} \\\hline
\end{tabular} 
\end{center}
\end{table}

For an ideal observer, $\eta_{p} = 1$. The lower the value of $\eta_{p}$ is, the slower the learning becomes, because the subject would require more data for similar updates in beliefs. If $\eta_{p} = 0$, the subject never learns from observing consecutive states. Note that the same parameter $\eta_{p}$ is used for learning all transition probabilities. 

\subsubsection*{\emph{\textbf{Implementation notes}}}
For prior beliefs about state transitions, a uniform prior would ensure that the prior does not privilege any probability value over another probability value. Then, for any entry $\alpha_{1}(s,s^\prime)$ of $\bm{\alpha}_{1}(s,s^\prime)$, we assume that $\alpha_{1}(s,s^\prime) = 1$ 

So, at spatial step $k$, for entry $\alpha_{s^\prime}(s,k)$ of $\bm{\alpha}(s,k)$,
\begin{align}\label{eq:p_stepk_rule}
    \alpha_k(s,s^\prime) = \eta_p c_{(s,s^\prime)}(k) + 1.
\end{align}

where $c_{(s,s^\prime)}(k)$ is the total number of observed transitions from $s$ to $s^\prime$ from step $1$ to step $k$. By keeping track of $c_{(s,s^\prime)}(k)$ in a matrix, any parameter in $\mathbf{A}(k)$ can be calculated on demand using Equation \ref{eq:p_stepk_rule} for the current state.

\chapter{Behavioral model part 2: the generative model}\label{chap:model_part2}

In the previous chapter, I discussed the internal representations of spatial regions and reward probabilities within those regions. This chapter describes a model that utilizes internal representations to generate behaviour. The learning model for updating beliefs about reward probabilities and state transitions utilized a normative model of Bayesian learning. In contrast, we present a descriptive model of behaviour that does not explicitly enforce any optimal decision-making criteria. Before making normative assumptions about behaviour, it is important to have a descriptive framework for systematically assessing assumptions about behaviour. 

Recall that location, visual stimulus, licking and speed of the mouse are recorded in the experimental data (see Chapter \ref{sec:data_description}). To improve readability, Table \ref{tab:discrete_behavioral_data2} includes notation used to represent the behavioural data.

\begin{table} [ht!]
\begin{center}
\captionsetup{justification=raggedright,singlelinecheck=false}
\caption{Behavioral and observational records for $t \in\{ 1, 2 \ldots, N\}$. }\label{tab:discrete_behavioral_data2} 
\begin{tabular}{|c|c|l|} \hline
     \textbf{Data} & Type &\multicolumn{1}{|c|}{\textbf{Description}} \\\hline
    $x_t$ & Observation &\specialcell{$x_t$ is the true value of the distance from the onset of the current \\ corridor at time step $t$. }  \\ \hline
    $y_t$ & Observation & \specialcell{$y_t \in Cor = \{ grey, vertical, angled \}$ is the true value of the \\corridor type, which determines the visual stimuli at time\\ step $t$. } \\ \hline
    $o_t$ & Observation &\specialcell{$o_t$ is a binary value for whether the reward valve has opened\\ during the time step. }  \\ \hline
    $v_t$ & Behavior &\specialcell{Speed (average) at time step $t$}  \\ \hline
    $l_t$ & Behavior & Number of licks at time step $t$ \\ \hline
\end{tabular}
\end{center}
\end{table}

\begin{figure}[hb!]
\centering
\includegraphics[scale=0.9]{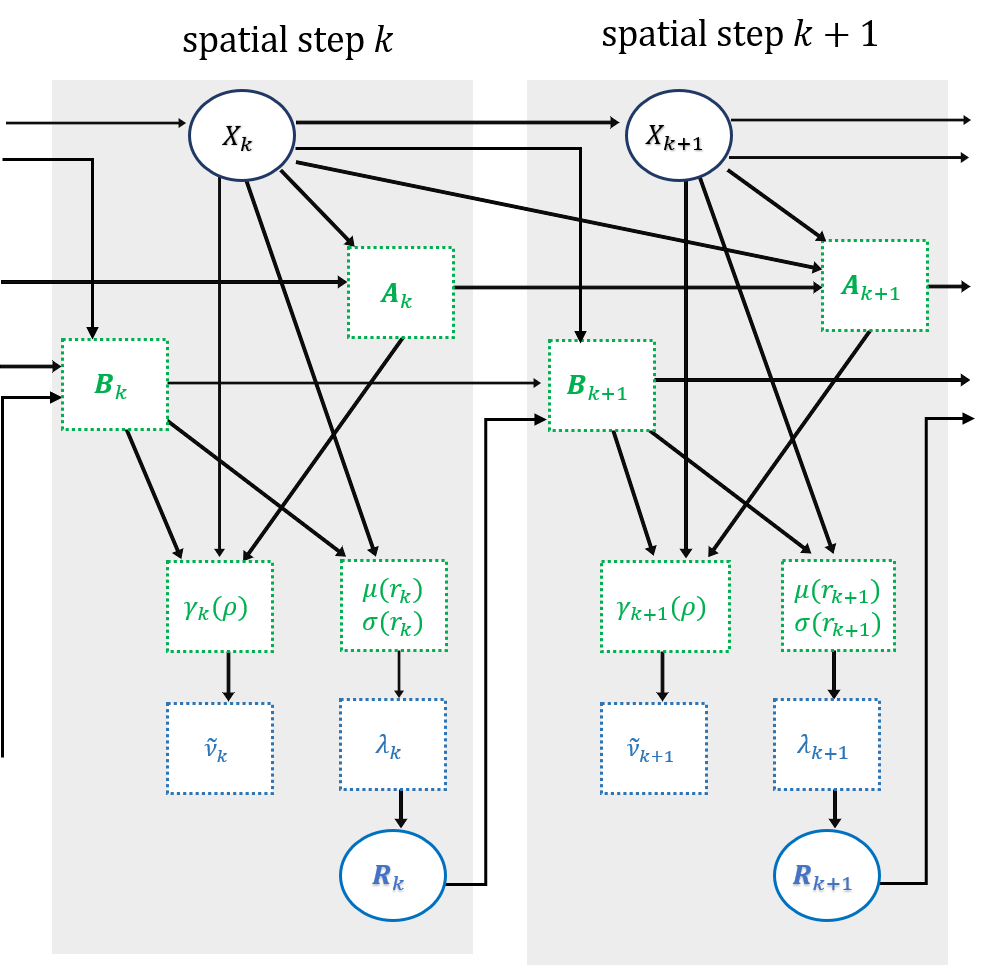}
\captionsetup{justification=raggedright,singlelinecheck=false}
\caption{Graphical model of updating internal representations at a given spatial step, the associated learning parameters (green), and the associated behavioural parameters (blue). The dotted squares indicate internal representations that are not observed in the data. Variables inside circles have stochastic outcomes given their parents, and variables inside squares have deterministic outcomes given their parents. State transitions trigger updating these variables for the new step $k+1$. Note that the model satisfies the Markov property. A description of the conditional dependencies is included in Table \ref{tab:learning_model_update}.}
\label{fig:steps_internal_reps}
\end{figure}

A spatial state transition event triggers updating internal representations of reward probability and spatial transitions. During the period between two transition events, the parameters associated with internal representations (specified by elements of $\bm{B}_k$ and $\bm{A}_k$) are unchanged. Assuming that the internal representations are guiding the behaviour, we define behavioural parameters for speed and licking rate derived from internal representations' parameters. Figure \ref{fig:steps_internal_reps} describes the conditional dependence structure of parameters associated with a spatial state. In this model, the internal representations are used to derive two parameters that guide the licking and speed behaviour. These parameters are \emph{target speed} $\tilde{\nu_k}$, and \emph{licking rate} $\tilde{\lambda_k}$, and they are discussed in detail in the Section \ref{sec:target_speed} and Section \ref{sec:licking_rate} respectively. 

\begin{table} [ht!]
\begin{center}
\captionsetup{justification=raggedright,singlelinecheck=false}
\caption{Description of updating internal representations of a given step using the graphical model of \ref{fig:steps_internal_reps}. Variables (\emph{Var.}) and their parents \emph{(Par(.))} are included in the first and second columns respectively. The third column \emph{(Type)} indicates whether the outcome of the variable given its parents is stochastic (\emph{Stoch.}) or deterministic (\emph{Deter.}) given its parents. The conditional dependence of the variable on its parents is described in the last column. }\label{tab:learning_model_update} 
\begin{tabular}{|c|c|c|l|} \hline
     \textbf{Var.} & \textbf{Par(.)} & \textbf{Type} &\multicolumn{1}{|c|}{\textbf{Update description}} \\\hline
   $X_{k+1}$ & \specialcell{$X_{k}$} & Stoch. &\specialcell{Stochastic outcome of the state immediately \\following $X_{k}$. }\\ \hline
   $\bm{B}_{k+1}$ & \specialcell{$\bm{B}_{k},\bm{R}_k,X_k$} & Deter. &\specialcell{Updating reward probability distribution of the \\previous state using Equation \ref{eq:r_update_rule}. \\}\\ \hline
   $\bm{A}_{k+1}$ & \specialcell{$\bm{A}_k, X_{k+1}, X_k$\\} & Deter. &\specialcell{
   Updating the transition probability distribution\\ for the last transition using Equation \ref{eq:p_update_rule}.\\
   distributions for reward, to $\bm{B}_{k+1}$}\\ \hline
   $r_{k}$ & $\bm{B}_{k},X_k$ & Deter. &\specialcell{Reward distribution of the current state}\\ \hline
   $\gamma_{k}(\rho)$ & $\bm{B}_{k},\bm{A}_{k}, \bm{X}_{k}$ & Deter. &\specialcell{ Discounted reward probability of present and \\future states given by Equation \ref{eq:discounted_reward}, with the \\ discount factor $\rho$.}\\ \hline
   $\tilde{\nu}_k$ & $\gamma_{k}(\rho)$ & Deter. &\specialcell{Value of target speed in spatial step $k$ adjusted by\\ value of $\gamma_{k}(\rho)$.\\
                                     }\\ \hline
   $\tilde{\lambda}_{k}$ & $r_{k}$ & Deter. &\specialcell{Licking rate in step $k$ given by Equation \ref{eq:lick_rate}.
                                     }\\ \hline
   $R_{k}$ & $\tilde{\lambda}_{k}$ & Stoch. &\specialcell{Reward outcome of spatial state $k$
                                     }\\ \hline

\end{tabular}
\end{center}
\end{table}

\section{Spatial state parameter \texorpdfstring{$\tilde{\lambda_k}$}{Lg}: licking rate} \label{sec:licking_rate}
Consider the relevance of the reward probability distribution for $r_k$ to the licking behaviour. First, it is reasonable to consider the mouse regulating its licking rate using its perception of expected reward probability in the current state. The expected value of the reward probability in the current state (in step $k$) is the expected value of $Beta(\beta^{(1)}_k(s),\beta^{(2)}_k(s))$, which is
\begin{align}\label{eq:average_r}
\mu(r_k) = \frac{\beta^{(1)}_k}{\beta^{(1)}_k \beta^{(2)}_k}.
\end{align}
Second, independently from the expectation of reward, the degree of uncertainty about the true probability of reward may also be relevant to behaviour \citep{zhao2015you}, and in particular, the rate of licking in the current state. More variance in the reward probability may mean that the current state should be further explored by licking, to decrease the uncertainty about reward values. The variance reward probability beliefs can also be calculated from the $Beta()$ distribution.
\begin{align}\label{eq:average_std}
\sigma^2(r_k) = \frac{\beta^{(1)}_k\beta^{(2)}_k}{(\beta^{(1)}_k+ \beta^{(2)}_k)^2 ~(\beta^{(1)}_k+ \beta^{(2)}_k+1)}
\end{align}
Let $L_t$ be a random variable for the number of licks at time step $t$. We assume that the licking rate is generated by a Poisson distribution 
\begin{align*}
    L_t \sim Pois(\tilde{\lambda_k})
\end{align*}
where for model parameters $\omega_1$, $\omega_2$ and $\omega_3$,
\begin{align}\label{eq:lick_rate}
    \tilde{\lambda_k} = \omega_1 \mu(r_k) + \omega_2 \sigma(r_k) + \omega_3,
\end{align}
is the licking rate at a time step spent within the current spatial step. The probability that $L_t = l_t$, for a number of licks $l_t$ is given by
\begin{align}\label{eq:lick_dist}
    P(L_t = l_t) = \frac{\lambda_k^{l_t} e^{-\lambda_k}}{l_t!}
\end{align}
\begin{table} [b]
\begin{center}
\caption{Parameters relevant to the licking behaviour. } \label{tab:licking_pars}.
\begin{tabular}{|c|l|l|} \hline
     \textbf{Parameter} & \textbf{Type} & \textbf{Description}  \\\hline
     {$\tilde{\lambda_k}$} & \specialcell{Spatial state\\ parameter}& \specialcell{Rate of the Poisson distribution generating the \\licking behavior within a time step spent in spatial\\ step $k$} \\ \hline
     $\omega_{1}$ & Model parameter & \specialcell{Weight of the expected reward probability of the\\ current reward distribution for calculating the\\ spatial state parameter $\tilde{\lambda_k}$ }  \\ \hline
     $\omega_{2}$ & Model parameter & \specialcell{Weight of the standard deviation of the current\\ reward distribution for calculating the\\ spatial state parameter $\tilde{\lambda_k}$ }\\ \hline
     $\omega_{3}$ & Model parameter & \specialcell{base licking rate for calculating $\tilde{\lambda_k}$ } \\\hline
\end{tabular} 
\end{center}
\end{table}

\section{Parameter \texorpdfstring{$\tilde{\nu_k}$}{Lg}: target speed within the current spatial state } \label{sec:target_speed}
We noticed that the mouse tends to speed up if it does not expect a reward in upcoming states (for example, see Figures \ref{fig:speed_M31_main} and \ref{fig:speed_M70_main}). We model this behavior using a discounted measure of future rewards.  

\subsubsection*{\emph{\textbf{Discounted future reward}}}
Expected average reward probability $m$ steps after the current state $s$ can be formulated as follows
\begin{align}\label{eq:expected_future_r_in_k+m}
\sum_{s^\prime \in S} E(r[s^\prime]) P(X_{k+m} = s^\prime|X_k = s)
\end{align}

Value of $P(X_{k+m}|X_k)$ can be estimated by the transition probability matrix obtained by the expected value of transition probabilities and standard Markov chain transition properties (Equation \ref{eq:MC_Pm+k}) \citep{haggstrom2002finite}. To estimate the values of the transition probability matrix, we use the expected value of transition probability for $p_{(s,s^\prime)}$, using parameters of Dirichlet distributions for transition probabilities in $\bm{A_k}$;
\begin{align*}
    E[p_{(s,s^\prime)}] = \frac{\alpha_k(s,s^\prime)}{\mathlarger{\sum}_{s^{\prime\prime} \in Adj(s)} \alpha_k(s,s^{\prime\prime})}
\end{align*}
is the estimated probability value for $p_{(s,s^\prime)}$ entry of the transition probability matrix. To conclude the discussion for the calculation of expression \ref{eq:expected_future_r_in_k+m}, note that $E(r[s^\prime]) = \beta^{(1)}_k(s^\prime)/\big(\beta^{(1)}_k(s^\prime) \beta^{(2)}_k(s^\prime)\big).$

Now, let us define the \emph{discounted future reward $\gamma_{k}(\rho)$} for a fixed value of $\rho$ in the current step $k$ to be  
\begin{align}\label{eq:discounted_reward}
\gamma_{k}(\rho) \coloneqq  \frac{\mathlarger{\sum}^{\infty}_{m=0} \rho^m \mathlarger{\sum}_{s^\prime \in S} \Big(E[r(s^\prime)] P(X_{k+m}|X_k)\Big)}{\sum^{\infty}_{m=0} \rho^m}
\end{align}
Note that $\gamma_{k}(\rho)$ is a normalised sum of discounted present and future expected reward probability values. Similar to the value function in reinforcement learning \citep{sutton2018reinforcement}, or the concept of discounted cash flow in financial asset valuation \citep{damodaran2012investment}, it incorporates all future reward values by iteratively giving less weight to future rewards that are further away.    

When transitioning from one state to another, lower discounted future reward $\gamma_{k}(\rho)$ is likely to indicate that the next reward is further away. In this case, the mouse may choose to adjust its behavior \citep{kleinfeld2006active}, by speeding up to pass the unrewarded regions more quickly. Since the discounted value of future reward does not change as long as the mouse is in the same spatial state, the desired speed at the current spatial step can be modeled as a spatial state parameter. Let the \emph{target speed} $\tilde{\nu_k}$ for the current state be
\begin{align}\label{eq:target_speed}
\tilde{\nu_k} := v_{max} \Big(1 - \gamma_{k}(\rho)\Big)
\end{align}
where $v_{max}$ is a model parameter that puts an upper bound on the target speed. A simple model of speed for time step $t$ is the following
\vspace{-0.3cm}\begin{align}\label{eq:speed1}
v_{t} \sim \mathcal{N}(\tilde{\nu_k}, \sigma_{\nu}^2).
\end{align}
However, physical constraints on the movement does not permit an instant jump in speed when the spatial state changes. The alternative model of speed that takes the physical constraints into considerations (by adding more parameters), is
\vspace{-0.3cm}\begin{align}\label{eq:speed2}
v_{t+1} \sim \mathcal{N}(E[v_{t+1}],  Var[v_{t+1}]),
\end{align}
\vspace{-0.3cm}where, \begin{align}
(E[v_{t+1}],  Var[v_{t+1}]) = \begin{cases}
 (v_t + \delta_{v}^{\plus}, \sigma_{v}^2) & \text{       for } v_t < \tilde{\nu_k} - \epsilon,\\
 (v_t + \delta_{v}^{\minus}, \sigma_{v}^2) & \text{       for } v_t > \tilde{\nu_k} + \epsilon,\\
 (v_t, \sigma_{v}^2) & \text{       otherwise; i.e., for } v_t \in [\tilde{\nu_k} - \epsilon, \tilde{\nu_k} + \epsilon].\\
\end{cases}
\end{align}
where the model parameters $\delta_{v}^{\plus}$ and $\delta_{v}^{\minus}$ are constant values for acceleration and deceleration, $\sigma_{v}^2$ is the variance of speed outcome in the next time-step. Furthermore, model parameter $\epsilon$ determines the range where non-random acceleration or deceleration is not enforced. 
\begin{table} [t]
\begin{center}
\caption{Parameters relevant to the speed behavior. } \label{tab:speed_pars}.
\begin{tabular}{|c|l|l|} \hline
     \textbf{Parameter} & \textbf{Type} & \textbf{Description}  \\\hline
     $\rho$ & Model parameter & \specialcell{ Discount rate of future reward (Expression \ref{eq:discounted_reward})} \\\hline
     {$\tilde{\nu_k}$} & \specialcell{Spatial state\\ parameter}& \specialcell{ Target speed (Expression \ref{eq:target_speed})\\  } \\ \hline
     {$\sigma_{\tilde{\nu}}^2$} & \specialcell{Model parameter}& \specialcell{ Variance of speed in the first model (Expression \ref{eq:speed1})\\  } \\ \hline
     {$\sigma_{v}^2$} & \specialcell{Model parameter}& \specialcell{ Variance of speed change\\ Expression \ref{eq:speed2} (second model)\\  } \\ \hline
     $\delta_{v}^{\plus},\delta_{v}^{\minus}$ & Model parameter & \specialcell{Acceleration and deceleration rate (second model) }  \\ \hline
     $\epsilon$ & Model parameter & \specialcell{Range of random only of speed change (second model)}\\ \hline
\end{tabular} 
\end{center}
\end{table}

\section{Generative model of licking and speed} \label{sec:generativeModel}

For given spatial states structure (by fixing parameters $\mathcal{V}$ and $d$), there exists a function $f_{\mathcal{V},d}: (xLoc \times Cor) \to S$ that associates each position to states. Then it is possible to determine time steps associated with state transitions. In Chapter \ref{sec:state_model}, we assumed that the states are fully observable to the subject. Therefore, the subject knows the  value of $f_{\mathcal{V},d}$ at any current time step. 
\subsubsection*{\emph{Binary variable $K_t$: indicator of spatial state transition event}}
For the current time step $t$, let $K_t$ be a binary variable such that
\vspace{-0.3cm}\begin{align}\label{eq:state_transition_indicator}
K_{t+1} = \begin{cases}
    0, & \text{          for } f_{\mathcal{V},d}(x_t, y_t) = f_{\mathcal{V},d}(x_{t+1}, y_{t+1}) \\
    1, & \text{          for } f_{\mathcal{V},d}(x_t, y_t) \not= f_{\mathcal{V},d}(x_{t+1}, y_{t+1}). \\
  \end{cases}
\end{align}
That is to say, $K_t = 1$ if $(x_t, y_t)$ and $(x_{t+1}, y_{t+1})$ are not in the same state, ans so a state transition has occurred. Note that a spatial state transition triggers an update in the beliefs about the environment (reward probability within states and state transitions). Then the internal representations in the graphical model of Figure \ref{fig:steps_internal_reps} are updated to the next spatial step, and the behavioral parameters $\lambda_{k_{t+1}}$ and $\tilde{nu}_{k_{t+1}}$ correspond to the new spatial step. For $K_t = 0$, the behavioral parameters $\tilde{\lambda}_{k_{t+1}}$ and $\tilde{nu}_{k_{t+1}}$ remain unchanged from the previous time-step. 

\begin{figure}[hb!]
\centering
\includegraphics[scale=0.68]{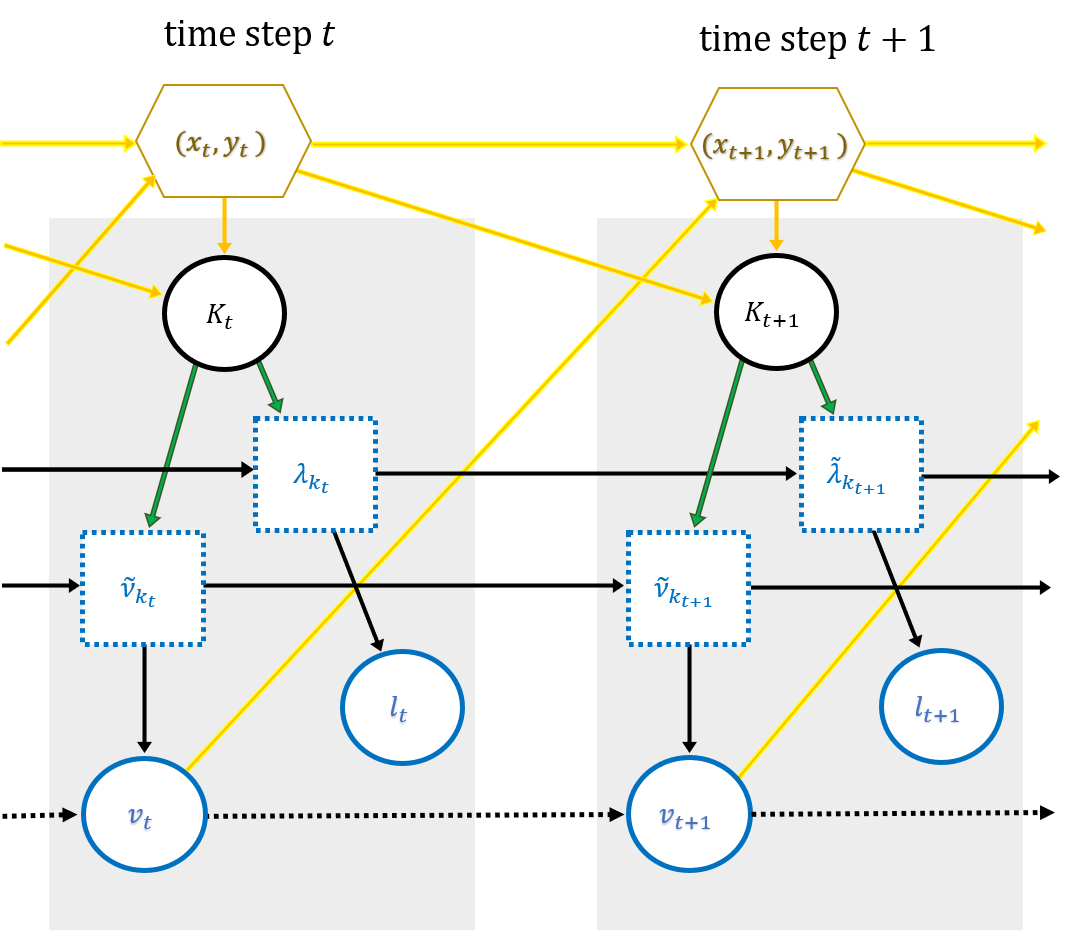}
\captionsetup{justification=raggedright,singlelinecheck=false}
\caption{Graphical model of the generative model of behavior. Note that the variables and relationships drawn in yellow and brown are not part of the internal model, and they describe the conditional dependence of the observed values to the model variables. See table \ref{tab:generative_model} for description of the relationships. }
\label{fig:generative_model}
\end{figure}

Figure \ref{fig:generative_model} is the graphical model for the generative model of behavior within time steps. The model assumes that the spatial state associated with $(x_t,y_t)$ is unambiguously determined by the subject (fully observable spatial states). Therefore, the value of $K_{t+1}$, which indicates a state transition, is also observed by the subject. Furthermore, $K_{t+1}$ can be deterministically inferred from the experimental data using the Equation \ref{eq:state_transition_indicator}. Hence, it is also observed in the behavioral data. If $K_{t+1} = 1$, then the graphical model of updating internal representations is used to find the new behavioral parameters (indicated by green arrows). If $K_{t+1} = 0$, the behavioral parameters remain unchanged from the previous step. A description of the relationships is included in Table \ref{tab:generative_model}.

\begin{table} [t!]
\begin{center}
\captionsetup{justification=raggedright,singlelinecheck=false}
\caption{Description of relationships in the  generative model of behavior in the graphical model of \ref{fig:generative_model}. Variables (\emph{Var.}) and their parents \emph{(Par(.))} are included in the first and second column respectively. Third column \emph{(Type)} indicates whether the outcome of the variable given its parents is stochastic (\emph{Stoch.}) or deterministic (\emph{Deter.}) given its parents. The conditional dependence of the variable on its parents is described in the last column. }\label{tab:generative_model} 
\begin{tabular}{|c|c|c|l|} \hline
     \textbf{Var.} & \textbf{Par(.)} & \textbf{Type} &\multicolumn{1}{|c|}{\textbf{Update description}} \\\hline
   $K_{t}$ & \specialcell{$(x_{t}, y_{t})$\\$(x_{t+1}, y_{t+1})$} & Stoch. &\specialcell{Transition event indicator (Expression \ref{eq:state_transition_indicator}). }\\ \hline
   $\tilde{\nu}_{k_{t+1}}$ & \specialcell{$\tilde{\nu}_{k_{t}},K_{t}$} & Deter. &\specialcell{\vspace{0.2cm}For $K_{t} = 0$, $\tilde{\nu}_{k_{t+1}}=\tilde{\nu}_{k_{t}}$. Otherwise, spatial state changes,\\ and graphical model $\ref{fig:steps_internal_reps}$ updates the value. }\\ \hline
   $\tilde{\lambda}_{k_{t+1}}$ & \specialcell{$\tilde{\lambda}_{k_{t}},K_{t}$} & Deter. &\specialcell{For $K_{t} = 0$, $\tilde{\lambda}_{k_{t+1}}=\tilde{\lambda}_{k_{t}}$. Otherwise, spatial state changes,\\ and graphical model $\ref{fig:steps_internal_reps}$ updates the value.). }\\ \hline
    $l_{t}$ & $\tilde{\lambda_k}$ & Stoch. &\specialcell{Poisson distributed value with rate $\tilde{\lambda_k}$
    (Expression \ref{eq:lick_rate}) }\\ \hline
   $v_{t}$ & $ \tilde{\nu_k} $ & Stoch. &\specialcell{Speed at time step $t$ by first model (Expression \ref{eq:speed1} ), \\or second model (Expression \ref{eq:speed2}. }\\ \hline
\end{tabular}
\end{center}
\end{table}

\pagebreak \section{Estimation of model parameters} \label{sec:parameters_estimation}

Below, the general framework for estimating the model parameters is discussed. For a fixed spatial model of space $\mathcal{M}_{V,d}$, let $\bm{\theta}$ be the list of model parameters
\begin{align*}
    \bm{\theta} &\coloneqq (\mathcal{V}, d, \eta_r, \eta_p, \omega_1, \omega_2, \omega_3, \sigma^2_{\tilde{\nu}}), & &\mbox{(using the second speed model), or}\\
               \bm{\theta}  &\coloneqq (\mathcal{V}, d, \eta_r, \eta_p, \omega_1, \omega_2, \omega_3, \sigma^2_{\tilde{v}}, \delta_{v}^{\plus},\delta_{v}^{\minus}, \epsilon)  &&\mbox{(using the first speed model).}\
\end{align*}
Given the model parameters, and given observational data, parents of $v_t$ and $l_t$ are deterministically set at each time point (see graphical model \ref{fig:generative_model}). Therefore, speed and licking are independent. So model likelihood of the generative model of behaviour at time step $t$ is
\begin{align*}
    \mathcal{L}\big(\bm{\theta}|(v_t,l_t)\big) &= P(v_t,l_t|\bm{\theta}) = P(v_t|\bm{\theta}) \ P(l_t|\bm{\theta})\\
    &\sim f\big(\mu_t(\bm{\theta}),\sigma_t(\bm{\theta})\big)\ g\big(l_t; \lambda_t(\bm{\theta}) \big)
\end{align*}

where $f$ are $g$ are probability mass functions for Gaussian and Poisson distributions respectively. Note that their distribution parameters are deterministically fixed at each time point given the model parameters (see Equations \ref{eq:lick_rate}, \ref{eq:speed1} and \ref{eq:speed2}). Then \emph{model evidence} for the generative model for up to time step $N$ is 
\begin{align} \label{eq:model_evidence}
 \mathcal{L}\Big(\bm{\theta}\Big|\{(v_t,l_t):t = 1 \ldots N\}\Big) \propto \prod_{t = 1}^{N} f\big(v_t; \mu_t(\bm{\theta}),\sigma_t(\bm{\theta})\big)\ g\big(l_t; \lambda_t(\bm{\theta}) \big)
\end{align}

And we can then use the maximum likelihood estimation (MLE) to estimate the fitted model parameters 
\begin{align}\label{eq:log_likelihood}
\bm{\bm{\theta}}^* =  \underset{\bm{\theta}}{\operatorname{argmax}} \sum_{t = 1} ^ N ln\Big[f\big(v_t; \mu_t(\bm{\theta}),\sigma_t(\bm{\theta})\big)\ g\big(l_t; \lambda_t(\bm{\theta}) \big)\Big]
\end{align}

Note that for each spatial step, the graphical model is used for calculating the parameters $\mu_t(\bm{\theta}),\sigma_t(\bm{\theta})$ and $\lambda_t(\bm{\theta})$.

\chapter{Discussion}\label{chap:discussion}

The next step in the project is to first complete the model validation on synthetic data. Before applying the model to real data, it is important to scrutinize the behaviour of the generative model. We plan to do so by pre-determining values for a model parameter and generating synthetic behavioural data. The generated behaviour is then used as a given data set. If the model is well-behaved, the model parameters should be recoverable from the synthetic data. As different spatial state structures radically alter the learning dynamics, we will conduct the parameter recovery for spatial model parameters more diligently. By considering various alternative hypotheses (different values for $d$ and $\mathcal{V}$), the model evidence (equation \ref{eq:model_evidence}) of alternative hypotheses will be compared. For a well-behaved model, the model evidence for the parameters used to generate data is expected to be the best.

\section{Limitations}

While our model assumes fully observable Markov states, noisy observations of the location and visual stimuli introduce uncertainty about the true current state of the system. Indeed, observations of the environment are often noisy and some behavioural models take this into account \citep{kangeye,kersten2009ideal}. While the learning rates of reward probability and transition probability capture some aspects of noisy observations, they are not based on normative assumptions. Alternatives should be considered for future research \citep{laquitaine2018switching}. Fortunately, there is an extensive body of research on partially observable Markov decision processes \citep{monahan1982state,kaelbling1996reinforcement} that would provide a clear path for improving the current model. 

An alternative to estimating the model parameters using MLE in Chapter \ref{sec:parameters_estimation} is to use the maximum a posteriori estimation (MAP) \citep{murphy2012machine,griffiths2008primer}. In contrast to MLE, which gives one estimated value for each parameter, MAP gives a distribution for each parameter, characterising the level of uncertainty about each parameter. Since some of the model parameters are qualitatively interpretable, MAP may be particularly relevant. In particular, a distribution over possible options for $\mathcal{V}$, the set of discriminated visual stimuli, is highly relevant to the imaged activity of the visual cortex. The potential challenge of MAP is that the computational difficulty of the calculation may introduce implementation challenges that are difficult to resolve. Nonetheless, its estimation of model parameters are potentially more meaningful for studying visual perception. 

\section{Implications}

During the experiments, two-photon calcium imaging and optogenetics were performed to determine changes in inputs and activity of individual excitatory and inhibitory cells within the primary visual cortex. Previously, a multivariate auto-regressive linear model (MVAR) was fitted to the neuronal data \citep{khan2018}:
\begin{align*}
    \bm{q}_{t+1} =  \bm{q}_t + A \times \bm{q}_t + \bm{u}_t + \xi v_t
\end{align*}
where $\bm{q}_t$ is the vector of response levels at time step $t$ for all $n$ imaged neurons, $A$ is an $n\times n$ matrix that includes the fitted interaction parameters, $\bm{u}_t$ is a fitted vector for the stimulus-related input, and $\xi$ is a fitted parameter for the contribution of current speed $v_t$. The MVAR model was used to compare the activity of populations of different inhibitory and excitatory cell types. The only behavioural term that was included was speed $v_t$, which did not make a significant contribution. An immediate application of the current behavioural model presented in this report is to potentially improve the MVAR model by including parameters related to internal representations, In particular, learned parameters that are likely to be relevant to behaviour, namely expected reward probability $\mu(r_k)$, variance $\sigma^2(r_k)$, and discounted future reward $\gamma_k(\rho)$ could potentially improve the predictive power of the MVAR model. 

If the internal representation terms from the behavioural model improve the predictive power of the MVAR model, it will give new insights into the information encoded in neurons located in the primary visual cortex. Future experiments can then be designed to systematically manipulate these internal terms to understand the precise representations \citep{heilbron2020word}. This will help us understand how the structure of the environment changes learning dynamics and internal representations.
\pagebreak


\bibliographystyle{agsm}
\bibliography{99-references}

\end{document}